\documentclass[12pt]{article}
\usepackage{epsf}
\title{ {\bf Lepton flavor violating $Z\rightarrow l_1^+ l_2^-$ decays with
the localized new Higgs doublet in the extra dimension}}
\author{\vspace{1cm}\\
        {\bf E. O. Iltan}
        \thanks{E-mail address:
        eiltan@heraklit.physics.metu.edu.tr}
 \\
        Physics Department, Middle East Technical University \\
        Ankara, Turkey\\}

\date{}

\begin{document}
\setlength{\baselineskip}{24pt}
\maketitle
\setlength{\baselineskip}{7mm}
\begin{abstract}
We predict the branching ratios of $Z\rightarrow e^{\pm}
\mu^{\pm}$, $Z\rightarrow e^{\pm} \tau^{\pm}$ and $Z\rightarrow
\mu^{\pm} \tau^{\pm}$ decays in the framework of the 2HDM with the
inclusion of one and two extra dimensions, by considering that the
new Higgs doublet is localized in the extra dimension with a
Gaussian profile. We observe that their BRs  are at the order of
the magnitude of $10^{-10}$, $10^{-8}$ and $10^{-5}$ with the
inclusion of a single extra dimension, in the given range of the
free parameters. These numerical values are slightly suppressed in
the case that the localization points of new Higgs scalars are
different than origin.
\end{abstract}
\thispagestyle{empty}
\newpage
\setcounter{page}{1}
\section{Introduction}
The lepton flavor violating (LFV) interactions are sensitive t0
the physics beyond the standard model (SM) and they are rich
theoretically since they exist at least in the one loop level. The
Z decays with different lepton flavor outputs, such as
$Z\rightarrow e \mu$, $Z\rightarrow e \tau$ and $Z\rightarrow \mu
\tau$ are among the candidates of LFV decays and they are clean in
the sense that they are free from the long distance effects. In
the literature, there is an extensive work on these decays
\cite{Riemann}-\cite{Perez}. The theoretical studies on  such Z
decays have been stimulated by the Giga-Z option of the Tesla
project which aims to increase the production of Z bosons at
resonance.

In the framework of the SM the lepton flavor is conserved and its
extension with massive neutrinos, so called $\nu$SM model, permits
the LFV interactions with the lepton mixing mechanism
\cite{Pontecorvo}. However, in this model, the theoretical
predictions of the branching ratios (BRs) of these LFV Z decays
are extremely small when the internal light neutrinos are light
\cite{Riemann,Illana}
\begin{eqnarray}
BR(Z\rightarrow e^{\pm} \mu^{\pm})\sim BR(Z\rightarrow e^{\pm}
\tau^{\pm})
&\sim& 10^{-54}  \nonumber \, , \\
BR(Z\rightarrow \mu^{\pm} \tau^{\pm}) &<& 4\times 10^{-60} .
\label{Theo1}
\end{eqnarray}
These numbers are far from the experimental limits obtained at LEP
1 \cite{PartData}:
\begin{eqnarray}
BR(Z\rightarrow e^{\pm} \mu^{\pm}) &<& 1.7\times 10^{-6} \,\,\,
\cite{Opal}
\nonumber \, , \\
BR(Z\rightarrow e^{\pm} \tau^{\pm}) &<& 9.8\times 10^{-6}\,\,\,
\cite{Opal,L3} \nonumber \, , \\
BR(Z\rightarrow \mu^{\pm} \tau^{\pm}) &<& 1.2\times 10^{-5} \,\,\,
\cite{Opal,Delphi} \label{Expr1}
\end{eqnarray}
and from the improved ones at Giga-Z \cite{Wilson}:
\begin{eqnarray}
BR(Z\rightarrow e^{\pm} \mu^{\pm}) &<& 2\times 10^{-9}  \nonumber \, , \\
BR(Z\rightarrow e^{\pm} \tau^{\pm}) &<& f\times 6.5\times 10^{-8}
\nonumber \, , \\
BR(Z\rightarrow \mu^{\pm} \tau^{\pm}) &<& f\times 2.2\times
10^{-8} \label{Expr2}
\end{eqnarray}
with $f=0.2-1.0$. Here the BRs are obtained for the decays
$Z\rightarrow \bar{l}_1 l_2+ \bar{l}_2 l_1$, namely,
\begin{eqnarray}
BR(Z\rightarrow l_1^{\pm} l_2^{\pm})=\frac{\Gamma (Z\rightarrow
\bar{l}_1 l_2+ \bar{l}_2 l_1)}{\Gamma_Z} \, .
\end{eqnarray}

The extensions of $\nu$SM with one heavy ordinary Dirac neutrino
\cite{Illana}, and with two heavy right-handed singlet Majorana
neutrinos \cite{Illana} ensure to enhance the BRs of the
corresponding LFV Z decays. The possible enhancements in their BRs
have been obtained in other models; in the the Zee model
\cite{Ghosal}, the two Higgs doublet model (2HDM), without (with)
the inclusion of the extra dimension
\cite{EiltZl1l2}(\cite{EiltZl1l2Extr}), the supersymmetric models
\cite{Masip, Cao} and the top-color assisted technicolor model
\cite{Yue}.

In this work, we study the LFV processes $Z\rightarrow e^{\pm}
\mu^{\pm}$, $Z\rightarrow e^{\pm} \tau^{\pm}$ and $Z\rightarrow
\mu^{\pm} \tau^{\pm}$ in the framework of the 2HDM with the
inclusion of a single (two) extra dimension(s). Here the LFV
interactions are induced by the internal new neutral Higgs bosons
$h^0$ and $A^0$ at least in the one loop level. The extension of
the Higgs sector brings new contribution to the BRs of the
considered decays. On the other hand, the inclusion the extra
dimensions enhance the BRs since the particle spectrum extends
after the compactification of the extra dimensions.

The extra dimension idea was originated from the study of
Kaluza-Klein \cite{Kaluza} which was related to the unification of
electromagnetism and the gravity and the motivation increased with
the study on the string theory which was formulated in a
space-time of more than four dimensions. Since the extra
dimensions are hidden to the experiments at present, the most
favorable description is the compactification these new dimensions
to the surfaces with small radii. In the case of that the extra
dimensions are at the order of submilimeter distance, for two
extra dimensions, the hierarchy problem in the fundamental scales
could be solved and the true scale of quantum gravity would be no
more the Planck scale but in the order of electroweak (EW) scale
\cite{Dvali,ArkaniHamed}.

The effects of extra dimensions on various physical processes have
been studied in the literature extensively
\cite{Dvali}-\cite{LocalNewHiggsLFV}. In the extra dimension
scenarios, the compactification procedure leads to appear new
particles, namely Kaluza-Klein (KK) modes in the theory. If all
the fields feel the extra dimensions, so called universal extra
dimensions (UED), the extra dimensional momentum, therefore the KK
number at each vertex, is conserved. If some fields feel the extra
dimensions but not all in the theory, those extra dimensions are
called non-universal extra dimensions and this is the case where
the KK number at vertices is not conserved. The non-conservation
of the KK number at the vertex results in the possibility of the
existence of the tree level interaction of KK modes with the
ordinary particles. In another scenario, some fields are
considered to be localized in the extra dimension(s). In the split
fermion scenario \cite{Mirabelli}-\cite{Grossman}, the fermions
are assumed at different points in the extra dimension with
Gaussian profiles and this ensures a possible solution to the
hierarchy of fermion masses by considering the overlaps of fermion
wave functions in the extra dimensions. The localization of the
Higgs doublet in the extra dimension has been considered in
\cite{Surujon}, by introducing an additional localizer field. In
\cite{IltanLFVSplitFat} the branching ratios of the radiative LFV
decays have been studied in the split fermion scenario, with the
assumption that the new Higgs doublet is restricted to the 4D
brane or to a part of the bulk in one and two extra dimensions, in
the framework of the 2HDM. \cite{iltSplitHiggsLocal} is devoted to
analysis of the BRs of the radiative LFV decays in the case that
the new Higgs scalars were localized in the extra dimension with
the help of the localizer field and the SM Higgs was considered to
have a constant profile. In the recent work
\cite{LocalNewHiggsLFV}, the radiative LFV decays were studied
with the assumption that the new Higgs doublet was localized in
the extra dimension with a Gaussian profile, by an unknown
mechanism, however, the other particle zero modes have uniform
profile in the extra dimension.

The present work is devoted to the BRs of the LFV Z decays in the
2HDM, with the inclusion of one and two extra dimensions by
considering that the new Higgs doublet is localized in the extra
dimension with a Gaussian profile, by an unknown mechanism,
however, the other particle zero modes have uniform profile in the
extra dimension. First, we assume that the new Higgs doublet is
localized around origin and, second, we take the localization
point as different than the origin but near to that. We observe
that the BRs of the LFV Z decays $Z\rightarrow e^{\pm} \mu^{\pm}$,
$Z\rightarrow e^{\pm} \tau^{\pm}$ and $Z\rightarrow \mu^{\pm}
\tau^{\pm}$  reach to the values at the order of the magnitude of
$10^{-10}$, $10^{-8}$ and $10^{-5}$ with the inclusion of a single
extra dimension, in the given range of the free parameters. These
numerical values are slightly suppressed in the case that the
localization points of new Higgs scalars are different than
origin.

The paper is organized as follows: In Section 2, we present the
effective vertex and the BRs of LFV Z decays in the 2HDM with the
inclusion of extra dimensions. Section 3 is devoted to discussion
and our conclusions. In appendix section, we give the explicit
expressions of the factors appearing in the effective vertex.
\section{The effect of the localization of the new Higgs doublet
on the lepton flavor violating $Z\rightarrow l_1^+ l_2^-$ decays
in the framework of the two Higgs Doublet model.}
The LFV Z boson decays $Z\rightarrow l_1^- l_2^+$ exist at least
in the one loop level and, therefore, the theoretical values of
the BRs are extremely small in the SM. With the extension of the
Higgs sector in which the flavor changing neutral current (FCNC)
at tree level is permitted, there appear additional contributions
to the BRs of the LFV processes. The multi Higgs doublet models
are among the candidates for such models and, in the present work,
we consider the 2HDM with FCNC at tree level. The inclusion of the
the spatial extra dimension further enhances the BRs, since the
particle spectrum is extended after its compactification and the
KK modes of the fields which are accessible to the extra dimension
bring additional contributions. Here, the idea is to consider that
the new Higgs scalars are localized in the extra dimension, with
Gaussian profiles, by an unknown mechanism, and, the other
particles have constant zero mode profiles  in the extra
dimension.

The Yukawa Lagrangian responsible for the LFV interactions in a
single extra dimension reads,
\begin{eqnarray}
{\cal{L}}_{Y}=
\xi^{E}_{5\,\,ij} \bar{l}_{i L} \phi_{2} E_{j R} + h.c. \,\,\, ,
\label{lagrangian}
\end{eqnarray}
where $L$ and $R$ denote chiral projections, $L(R)=1/2(1\mp
\gamma_5)$, $\phi_{2}$ is the new scalar doublet and $\xi^{E}_{5\,
ij}$ are the FV Yukawa couplings in five dimensions, where $i,j$
are family indices of leptons, $l_{i}$ and $E_{j}$ are lepton
doublets and singlets respectively. These fields are the functions
of $x^\mu$ and $y$, where $y$ is the coordinate represents the
fifth dimension. Here we choose the Higgs doublets $\phi_{1}$ and
$\phi_{2}$ as
\begin{eqnarray}
\phi_{1}=\frac{1}{\sqrt{2}}\left[\left(\begin{array}{c c}
0\\v+H^{0}\end{array}\right)\; + \left(\begin{array}{c c} \sqrt{2}
\chi^{+}\\ i \chi^{0}\end{array}\right) \right]\, ;
\phi_{2}=\frac{1}{\sqrt{2}}\left(\begin{array}{c c} \sqrt{2}
H^{+}\\ H_1+i H_2 \end{array}\right) \,\, , \label{choice}
\end{eqnarray}
and their vacuum expectation values as
\begin{eqnarray}
<\phi_{1}>=\frac{1}{\sqrt{2}}\left(\begin{array}{c c}
0\\v\end{array}\right) \,  \, ; <\phi_{2}>=0 \,\, .
\label{choice2}
\end{eqnarray}
This choice makes it possible to collect the SM (new) particles in
the first (second) doublet and $H_1$ and $H_2$ becomes the mass
eigenstates $h^0$ and $A^0$, respectively since no mixing occurs
between two CP-even neutral bosons $H^0$ and $h^0$ at tree level.
%
%
%

The five dimensional lepton doublets and singlets and the SM Higgs
field are expanded into their KK modes with the compactification
of the extra dimension on an orbifold $S^1/Z_2$ with radius $R$
and they read
\begin{eqnarray}
\phi_{1}(x,y ) & = & {1 \over {\sqrt{2 \pi R}}} \left\{
\phi_{1}^{(0)}(x) + \sqrt{2}
\sum_{n=1}^{\infty}  \phi_{1}^{(n)}(x) \cos(ny/R)\right\}\, ,\nonumber\\
l_i (x,y )& = & {1 \over {\sqrt{2 \pi R}}} \left\{ l_{i
L}^{(0)}(x) + \sqrt{2} \sum_{n=1}^{\infty} \left[l_{i L}^{(n)}(x)
 \cos(ny/R) + l_{i R}^{(n)}(x) \sin(ny/R)\right]\right\}\, ,\nonumber\\
E_{i}(x,y )& = & {1 \over {\sqrt{2 \pi R}}} \left\{ E_{i
R}^{(0)}(x) + \sqrt{2} \sum_{n=1}^{\infty}  \left[E_{i R}^{(n)}(x)
\cos(ny/R) + E_{i L}^{(n)}(x) \sin(ny/R)\right]\right\} \,\, ,
\label{f0}
\end{eqnarray}
where $\phi_{1}^{(0)}(x)$, $l_{i L}^{(0)}(x)$ and $E_{i
R}^{(0)}(x)$ are the four dimensional Higgs doublet, lepton
doublets and lepton singlets respectively. On the other hand the
new Higgs scalar profiles read,
\begin{eqnarray}
S(x,y )=A e^{-\beta y^2} S(x) \label{phi2}\, ,
\end{eqnarray}
and the mechanism beyond the localization is unknown\footnote{We
consider the zero mode Higgs scalars and we do not take into
account the possible KK modes of Higgs scalars since the mechanism
for the localization is unknown and we expect that the those
contributions are small due to their heavy masses.}. Here the
normalization constant $A$ is
\begin{eqnarray}
A=\frac{(2\,
\beta)^{1/4}}{\pi^{1/4}\,\sqrt{Erf[\sqrt{2\,\beta}\,\pi\,R]}}\label{Norm}
\, .
\end{eqnarray}
The strength of the localization of the new Higgs doublet in the
extra dimension is regulated by the parameter $\beta=1/\sigma^2$,
where $\sigma$, $\sigma=\rho\,R$, is the Gaussian width of
$S(x,y)$ in the extra dimension. Here the function $Erf[z]$ is the
error function, which is defined as
\begin{eqnarray}
Erf[z]=\frac{2}{\sqrt{\pi}}\,\int_{0}^{z}\,e^{-t^2}\,dt \,\, .
\label{erffunc}
\end{eqnarray}
The modified Yukawa interactions in four dimensions can be
obtained by integrating the combination $\bar{f}^{(0
(n))}_{iL\,(R)}(x,y)\,S(x,y)\, f^{(n (0))}_{j R\,(L)}(x,y)$,
appearing in the Lagrangian (eq. (\ref{lagrangian})), over the
fifth dimension as
\begin{eqnarray}
I=\int_{-\pi R}^{\pi R}\, dy\,\, \bar{f}^{(0
(n))}_{iL\,(R)}(x,y)\,S(x,y)\, f^{(n (0))}_{jR\,(L)}(x,y)\,\, ,
\label{intVij11}
\end{eqnarray}
where $f^{(n (0))}_{j R\,(L)}$ are the KK basis (zero-mode) for
lepton fields (eq. (\ref{f0})), and finally, we get
\begin{eqnarray}
I=V_n \, \bar{f}^{(0(n))}_{iL\,(R)}(x) \,S(x)\,\,f^{(n (0))}_{j
R\,(L)}(x)\,\, , \label{intVij1}
\end{eqnarray}
with the factor $V_n$
\begin{eqnarray}
V_n=A \, c_n \, , \label{Vij1even}
\end{eqnarray}
and the function $A$ which is defined in eq. (\ref{Norm}). The
function $c_n$ in eq. (\ref{Vij1even}) is obtained as:
\begin{eqnarray}
c_ n=e^{-\frac{n^2}{4\,\beta\,R^2}}\, \frac{\Bigg( Erf[\frac{i\,
n+2\,\beta\,\pi\,R^2}{2\,\sqrt{\beta}\,R}]+Erf[\frac{-i\,
n+2\,\beta\,\pi\,R^2}{2\,\sqrt{\beta}\,R}]\Bigg)}
{4\,\sqrt{\beta\,\pi}\,R} \label{cevenodd} \, .
\end{eqnarray}
Notice that the factor $A$ is embedded into the definition of the
the Yukawa couplings $\xi^{E}_{ij}$ in four dimensions as
\begin{eqnarray}
\xi^{E}_{ij}= A \, \xi^{E}_{5\, ij}\, , \label{coupl4}
\end{eqnarray}
where $\xi^{E}_{5\, ij}$ are  the Yukawa couplings in five
dimensions (see eq. (\ref{lagrangian})).

In the following, we consider that the new Higgs scalars are
localized in the extra dimension at the point $y_H$,
$y_H=\alpha\,\sigma$ near to the origin, namely,
\begin{eqnarray}
S(x,y)=A_H \,e^{-\beta (y-y_H)^2}\, S(x) \label{phi2H}\, ,
\end{eqnarray}
with the normalization constant
\begin{eqnarray}
A_H=\frac{2\,( \beta)^{1/4}}{(2
\pi)^{1/4}\,\sqrt{Erf[\sqrt{2\,\beta}\,(\pi\,R+y_H)]+
Erf[\sqrt{2\,\beta}\,(\pi\,R-y_H)]}} \label{NormH} \, .
\end{eqnarray}
The integration of the combination $\bar{f}^{(0
(n))}_{iL\,(R)}(x,y)\,S(x,y)\, f^{(n (0))}_{j R\,(L)}(x,y)$ over
extra dimension brings the factor $V_n$ appearing in eq.
(\ref{intVij1}) as
\begin{eqnarray}
V_n=A_H \, c_n \, , \label{Vij1evenH}
\end{eqnarray}
where $A_H$ is the normalization constant defined in eq.
(\ref{NormH}) and  function $c_n$ reads
\begin{eqnarray}
c_ n=e^{-\frac{n^2}{4\,\beta\,R^2}}\, Cos[\frac{n\,y_H}{R}]\,
\frac{\Bigg( Erf[\frac{i\,
n+2\,\beta\,\pi\,R^2}{2\,\sqrt{\beta}\,R}]+Erf[\frac{-i\,
n+2\,\beta\,\pi\,R^2}{2\,\sqrt{\beta}\,R}]\Bigg)}
{4\,\sqrt{\beta\,\pi}\,R} \label{cevenoddH} \, .
\end{eqnarray}
Similar to the previous case, we define the Yukawa couplings in
four dimensions as
\begin{eqnarray}
\xi^{E}_{ij}= A_H \, \xi^{E}_{5\, ij}\, . \label{coupl4H}
\end{eqnarray}
Now, we would like to make the same analysis in  the case of two
spatial extra dimensions. The six dimensional lepton doublets and
singlets and the SM Higgs fields are expanded into their KK modes
with the compactification of the extra dimension on an orbifold
$(S^1\times S^1)/Z_2$, with radius $R$ and they read
\begin{eqnarray}
\phi_{1}(x,y,z) & = & {1 \over {2 \pi R}} \left\{
\phi_{1}^{(0,0)}(x) + 2 \sum_{n,s}  \phi_{1}^{(n,s)}(x)
\cos(ny/R+sz/R)\right\}\, , \nonumber\\ l_i (x,y,z)& = & {1 \over
{2 \pi R}} \left\{ l_{i L}^{(0,0)}(x) + 2 \sum_{n,s} \left[l_{i
L}^{(n,s)}(x) \cos(ny/R+sz/R) + l_{i R}^{(n,s)}(x)
\sin(ny/R+sz/R)\right] \right\}\, ,\nonumber\\ E_{i}(x,y,z)
\!\!\!\!& = & \!\!\!\!{1 \over {2 \pi R}} \left\{ \!\!E_{i
R}^{(0,0)}(x) + \!\! 2 \sum_{n,s} \left[E_{i R}^{(n,s)}(x)
\cos(ny/R+sz/R) + \!\! E_{i L}^{(n,s)}(x)
\sin(ny/R+sz/R)\right]\right\} \,\, , \nonumber\\ \label{f02}
\end{eqnarray}
where $\phi_{1}^{(0,0)}(x)$, $l_{i L}^{(0,0)}(x)$ and $E_{i
R}^{(0,0)}(x)$ are the four dimensional Higgs doublet, lepton
doublets and lepton singlets respectively. Here the summation is
done over the indices $n,s$ but both are not zero at the same
time. Similar to a single extra dimension case, the new Higgs
scalar profiles read,
\begin{eqnarray}
S(x,y,z )=A' e^{-\beta (y^2+z^2)} S(x) \label{phi22}\, ,
\end{eqnarray}
and the mechanism beyond its localization is unknown. Here
normalization constant $A'$ is
\begin{eqnarray}
A'=\frac{(2\,
\beta)^{1/2}}{\pi^{1/2}\,Erf[\sqrt{2\,\beta}\,\pi\,R]}\label{Norm2}
\, .
\end{eqnarray}
The modified Yukawa interactions in four dimensions can be
obtained by integrating the combination $\bar{f}^{(0,0
(n,s))}_{iL\,(R)}(x,y)\,S(x,y,z)\, f^{(n,s (0,0))}_{j
R\,(L)}(x,y)$ over the fifth and sixth dimensions:
\begin{eqnarray}
I=\int_{-\pi R}^{\pi R}\, dy\, \int_{-\pi R}^{\pi R}\,  dz\,\,
\bar{f}^{(0,0 (n,s))}_{iL\,(R)}(x,y,z)\,S(x,y,z)\, f^{(n,s
(0,0))}_{jR\,(L)}(x,y,z)\,\, , \label{intVij112}
\end{eqnarray}
where
\begin{eqnarray}
I=V_{n,s} \, \bar{f}^{(0,0(n,s))}_{iL\,(R)}(x) \,S(x)\,\,f^{(n,s
(0,0))}_{j R\,(L)}(x)\,\, , \label{intVij12}
\end{eqnarray}
with the factor $V_{n,s}$
\begin{eqnarray}
V_{n,s}=A' \, c_{n,s} \, , \label{Vij1even2}
\end{eqnarray}
and the function $A'$ which is defined in eq. (\ref{Norm2}). The
function $c_{n,s}$ in eq. (\ref{Vij1even2}) is obtained as:
\begin{eqnarray}
c_{n,s}=e^{-\frac{n^2+m^2}{4\,\beta\,R^2}}\, \frac{\Bigg(
Erf[\frac{i\,
n+2\,\beta\,\pi\,R^2}{2\,\sqrt{\beta}\,R}]+Erf[\frac{-i\,
n+2\,\beta\,\pi\,R^2}{2\,\sqrt{\beta}\,R}]\Bigg)\, \Bigg(
Erf[\frac{i\,
s+2\,\beta\,\pi\,R^2}{2\,\sqrt{\beta}\,R}]+Erf[\frac{-i\,
s+2\,\beta\,\pi\,R^2}{2\,\sqrt{\beta}\,R}]\Bigg)}
{16\,\beta\,\pi\,R^2} \label{cevenodd2} ,
\end{eqnarray}
Here the Yukawa couplings $\xi^{E}_{ij}$ in four dimensions read
\begin{eqnarray}
\xi^{E}_{ij}= A' \, \xi^{E}_{6\, ij}\, , \label{coupl42}
\end{eqnarray}
where $\xi^{E}_{6\, ij}$ are  the Yukawa couplings in six
dimensions.

The internal neutral Higgs particles $h^0$ and $A^0$ play the main
role in the existence of the $Z\rightarrow l_1^- l_2^+$ decay,
theoretically. In Fig. \ref{fig1ver} the necessary 1-loop
diagrams, the self energy and vertex diagrams, are given. The
inclusion of extra dimensions brings  additional lepton KK mode
contributions. The general effective vertex for the interaction of
on-shell Z-boson with a fermionic current reads
\begin{eqnarray}
\Gamma_{\mu}=\gamma_{\mu}(f_V-f_A\ \gamma_5)+
\frac{i}{m_W}\,(f_M+f_E\, \gamma_5)\, \sigma_{\mu\,\nu}\,
q^{\nu}\,  , \label{vertex}
\end{eqnarray}
where $q$ is the momentum transfer, $q^2=(p-p')^2$, $f_V$ ($f_A$)
is vector (axial-vector) coupling, $f_M$ ($f_E$) magnetic
(electric) transitions of unlike fermions. Here $p$
($-p^{\prime}$) is the four momentum vector of lepton
(anti-lepton). The  vector (axial-vector) $f_V$ ($f_A$) couplings
and the magnetic (electric) transitions $f_M$ ($f_E$) including
the contributions coming from a single extra dimension can be
obtained as
\begin{eqnarray}
f_V&=&\sum_{i=1}^{3} \Big( f_{iV}^{(0)}+2 \sum_{n=1}^{\infty}
f_{iV}^{(n)} \Big)
\nonumber \, , \\
f_A&=&\sum_{i=1}^{3} \Big( f_{iA}^{(0)}+2 \sum_{n=1}^{\infty}
f_{iA}^{(n)}
\Big)\nonumber \, , \\
f_M&=&\sum_{i=1}^{3} \Big( f_{iM}^{(0)}+2 \sum_{n=1}^{\infty}
f_{iM}^{(n)}
\Big)\nonumber \, , \\
f_E&=&\sum_{i=1}^{3} \Big( f_{iE}^{(0)}+2 \sum_{n=1}^{\infty}
f_{iE}^{(n)} \Big)\, , \label{fVAMEex}
\end{eqnarray}
where $f^{(0)}_{i(V,A,M,E)}$ are the couplings without lepton KK
mode contributions and they can be calculated by taking $n=0$ in
eq. (\ref{fVAME}). On the other hand the couplings
$f^{(n)}_{i(V,A,M,E)}$ are the ones due to the KK modes of the
leptons (see eq. (\ref{fVAME})). Here the summation over the index
$i$ represents the sum due to the internal lepton flavors, namely,
$e,\mu,\tau$. We present $f^{(n)}_{i(V,A,M,E)}$ in the appendix,
by taking into account all the masses of internal leptons and
external lepton (anti-lepton). If we consider two extra dimensions
where all the particles are accessible, the couplings
$f^{(n)}_{i(V,A,M,E)}$ appearing in eq. (\ref{fVAMEex}) should be
replaced by $f^{(n,s)}_{i(V,A,M,E)}$ and they read
\begin{eqnarray}
f_V&=&\sum_{i=1}^{3} \Big( f_{iV}^{(0,0)}+4 \sum_{n,s}^{\infty}
f_{iV}^{(n,s)} \Big)
\nonumber \, , \\
f_A&=&\sum_{i=1}^{3} \Big( f_{iA}^{(0,0)}+4 \sum_{n,s}^{\infty}
f_{iA}^{(n,s)}
\Big)\nonumber \, , \\
f_M&=&\sum_{i=1}^{3} \Big( f_{iM}^{(0,0)}+4 \sum_{n,s}^{\infty}
f_iM^{(n,s)}
\Big)\nonumber \, , \\
f_E&=&\sum_{i=1}^{3} \Big( f_{iE}^{(0,0)}+4 \sum_{n,s}^{\infty}
f_{iE}^{(n,s)} \Big)\, , \label{fVAMEex2}
\end{eqnarray}
where the summation would be done over $n,s=0,1,2 ...$ except
$n=s=0$ (see appendix for their explicit forms).

Finally, the BR for $Z\rightarrow l_1^-\, l_2^+$ can be written in
terms of the couplings $f_V$, $f_A$, $f_M$ and $f_E$ as
\begin{eqnarray}
BR (Z\rightarrow l_1^-\,l_2^+)=\frac{1}{48\,\pi}\,
\frac{m_Z}{\Gamma_Z}\,
\{|f_V|^2+|f_A|^2+\frac{1}{2\,cos^2\,\theta_W} (|f_M|^2+|f_E|^2)
\} \label{BR1}
\end{eqnarray}
where
$\Gamma_Z$ is the total
decay width of Z boson. In our numerical analysis  we consider the
BR due to the production of sum of charged states, namely
\begin{eqnarray}
BR (Z\rightarrow l_1^{\pm}\,l_2^{\pm})= \frac{\Gamma(Z\rightarrow
(\bar{l}_1\,l_2+\bar{l}_2\,l_1)}{\Gamma_Z} \, .\label{BR2}
\end{eqnarray}
%
\section{Discussion}
The LFV Z decays $Z\rightarrow l_1^{\pm} l_2^{\pm}$, $l_1\neq l_2$
are rare decays in the sense that they exist at least in the one
loop level and they are rich theoretically since the physical
parameters of these decays contain number of free parameters of
the model used. In the framework of the 2HDM the internal leptons
and new scalar bosons drive the interaction and the corresponding
physical quantities are sensitive to the Yukawa
couplings\footnote{In the following we use the dimensionful
coupling $\bar{\xi}^{E}_{N,ij}$ in four dimensions, with the
definition $\xi^{E}_{N,ij}=\sqrt{\frac{4\, G_F}{\sqrt{2}}}\,
\bar{\xi}^{E}_{N,ij}$ where N denotes the word "neutral".}
$\bar{\xi}^E_{N,ij}, \, i,j=e, \mu, \tau$, which are among the
free parameters of the model. These couplings should be restricted
by using present and forthcoming experiments. Here, we assume that
the couplings which contain $\tau$ index are dominant respecting
the Cheng-Sher scenario \cite{Sher} and, therefore, we consider
only the internal $\tau$ lepton in the loop diagrams. In addition
to this, we take the Yukawa couplings $\bar{\xi}^{D}_{N,ij}$ as
symmetric with respect to the indices $i$ and $j$. As a result,
among the Yukawa couplings we need the numerical values for the
$\bar{\xi}^{D}_{N,\tau e}$, $\bar{\xi}^{D}_{N,\tau \mu}$ and
$\bar{\xi}^{D}_{N,\tau \tau}$. Furthermore,  the new Higgs masses
are also free parameters of the model and we take their numerical
values as $m_{h^0}=100\, GeV$, $m_{A^0}=200\, GeV$.

In the present work, we study the LFV decays $Z\rightarrow
l_1^{\pm} l_2^{\pm}$, $l_1\neq l_2$ in the framework of the 2HDM
with the addition  extra dimensions. Our assumption is that the
new Higgs scalars are localized in the extra dimension with
Gaussian profiles by an unknown mechanism, however, the other
particles have uniform zero mode profiles in the extra dimension.
Here we choose one (two) extra dimension(s) which are compactified
on to orbifold $S^1/Z_2$ ($(S^1\times S^1)/Z_2$) with the
compactification scale $1/R$, which is another free parameter. The
direct limits from searching for KK gauge bosons imply $1/R>
800\,\, GeV$, the precision electro weak bounds on higher
dimensional operators generated by KK exchange place a far more
stringent limit $1/R> 3.0\,\, TeV$ \cite{Rizzo} and, from
$B\rightarrow \phi \, K_S$, the lower bounds for the scale $1/R$
have been obtained as $1/R > 1.0 \,\, TeV$, from $B\rightarrow
\psi \, K_S$ one got $1/R
> 500\,\, GeV$, and from the upper limit of the $BR$, $BR \, (B_s
\rightarrow \mu^+ \mu^-)< 2.6\,\times 10^{-6}$, the estimated
limit was $1/R > 800\,\, GeV$ \cite{Hewett}. On the other hand,
the localization of new Higgs doublet is regulated by the
parameter $\sigma$, which is the Gaussian width of new Higgs
doublet in the extra dimension, and it is chosen so that it does
not contradict with the experimental results.  Here, we take the
compactification scale $1/R$ in the range $200\, GeV\leq 1/R \leq
1000\, GeV$ and choose the Gaussian width $\sigma=\rho\, R $ at
most $0.05\,R$.
Notice that throughout our calculations we use the input values
given in Table (\ref{input}).
\newpage
\begin{table}[h]
        \begin{center}
        \begin{tabular}{|l|l|}
        \hline
        \multicolumn{1}{|c|}{Parameter} &
                \multicolumn{1}{|c|}{Value}     \\
        \hline \hline
        $m_{\mu}$                   & $0.106$ (GeV) \\
        $m_{\tau}$                   & $1.78$ (GeV) \\
        $m_{W}$             & $80.26$ (GeV) \\
        $m_{Z}$             & $91.19$ (GeV) \\
        $m_{h^0}$             & $100$ (GeV) \\
        $m_{A^0}$             & $200$ (GeV) \\
        $G_F$             & $1.16637 10^{-5} (GeV^{-2})$  \\
        $\Gamma_Z$                  & $2.490\, (GeV)$  \\
        $sin\,\theta_W$               & $\sqrt{0.2325}$ \\
        \hline
        \end{tabular}
        \end{center}
\caption{The values of the input parameters used in the numerical
          calculations.}
\label{input}
\end{table}

In our analysis, we first consider that the new Higgs doublet is
localized around the origin in a single extra dimension.
Furthermore, we choose the localization point is near to the
origin, at the point $y_H=\alpha \,\sigma$, and study its effect
on the BRs. We continue to analyze the same physical quantity with
the inclusion of two extra dimensions.

Fig. \ref{Zmuero} is devoted to the parameter $\rho=\sigma/R$
dependence of the BR $\,(Z\rightarrow \mu^{\pm}\, e^{\pm})$ for
$\bar{\xi}^{D}_{N,\tau e}=0.1\, GeV$, $\bar{\xi}^{D}_{N,\tau
\mu}=10\,GeV$ and $1/R=500\, GeV$. Here the solid (dashed, small
dashed) line represents the BR without lepton KK modes in a single
extra dimension (with lepton KK modes in a single extra dimension,
with lepton KK modes in two extra dimensions). It is observed that
BR is at the order of the magnitude of $10^{-14}$ without lepton
KK modes in a single extra dimension, for the the parameter
$\rho\sim 0.05$. In this case the BR is sensitive to $\rho$. With
the inclusion of lepton KK modes, the BR enhances  to the values
of the order of $10^{-10}$ and this is almost four order
enhancement in the BR. For two extra dimensions, the numerical
value of the BR is slightly smaller compared to the single extra
dimension case, since there is an additional suppression factor
(see the exponential factor in eq. (\ref{cevenodd2})) appears in
the expressions.

In Fig. \ref{Ztauero}, we present the $\rho$ dependence of the BR
$\,(Z\rightarrow \tau^{\pm}\, e^{\pm})$ for $\bar{\xi}^{D}_{N,\tau
e}=0.1\, GeV$, $\bar{\xi}^{D}_{N,\tau \tau}=100\,GeV$ and
$1/R=500\, GeV$. Here the solid (dashed, small dashed) line
represents the BR without lepton KK modes in a single extra
dimension (with lepton KK modes in a single extra dimension, with
lepton KK modes in two extra dimensions). This figure shows that
BR is at the order of the magnitude of $10^{-12}$ without lepton
KK modes in a single extra dimension, for the the parameter
$\rho\sim 0.05$ and its sensitivity to the parameter $\rho$ is
strong. The inclusion of lepton KK modes results in an
considerable enhancement in the BR almost four order and the BR of
the decay under consideration enhances to the values of the order
of $10^{-8}$. For two extra dimensions, the numerical value of the
BR is almost the same as the single extra dimension case.

Fig. \ref{Ztaumuro} represents the $\rho$ dependence of the BR
$\,(Z\rightarrow \tau^{\pm}\, \mu^{\pm})$ for
$\bar{\xi}^{D}_{N,\tau \mu}=10\, GeV$, $\bar{\xi}^{D}_{N,\tau
\tau}=100\,GeV$ and $1/R=500\, GeV$ . Here the solid (dashed,
small dashed) line represents the BR without lepton KK modes in a
single extra dimension (with lepton KK modes in a single extra
dimension, with lepton KK modes in two extra dimensions). We
observe that BR is at the order of the magnitude of $10^{-8}$
($10^{-5}$) without (with) lepton KK modes in a single extra
dimension, for the the parameter $\rho\sim 0.05$. Similar to the
previous decays, the inclusion of lepton KK modes results in an
considerable enhancement in the BR more than three orders. For two
extra dimensions, the numerical value of the BR is almost the same
as the single extra dimension case.

At this stage, we study the dependence of the BR of the LFV Z
decays to the Yukawa couplings, regulating the lepton-lepton-new
Higgs interactions. Figs.
\ref{Zmueksi}-\ref{Ztaueksi}-\ref{Ztaumuksi} represent the Yukawa
coupling $\bar{\xi}^{E}_{N,\tau e}$\,-\,$\bar{\xi}^{E}_{N,\tau
e}$\,-\,$\bar{\xi}^{E}_{N,\tau \tau}$ dependence of the BR
$\,(Z\rightarrow \mu^{\pm}\, e^{\pm})$\,-\,BR $\,(Z\rightarrow
\mu^{\pm}\, e^{\pm})$\,-\,BR $\,(Z\rightarrow \tau^{\pm}\,
\mu^{\pm})$ for $\bar{\xi}^{D}_{N,\tau \mu}=10\,
GeV$\,-\,$\bar{\xi}^{D}_{N,\tau \tau}=100\,
GeV$\,-\,$\bar{\xi}^{D}_{N,\tau \mu}=10\, GeV$, $\rho=0.01$ and
$1/R=500\, GeV$. Here the solid (dashed, small dashed) line
represents the BR without lepton KK modes in a single extra
dimension (with lepton KK modes in a single extra dimension, with
lepton KK modes in two extra dimensions) for all figures. The BR
is strongly sensitive to to the Yukawa coupling
$\bar{\xi}^{E}_{N,\tau e}$\,-\,$\bar{\xi}^{E}_{N,\tau
e}$\,-\,$\bar{\xi}^{E}_{N,\tau \tau}$ and in the interval $0.005
\leq \bar{\xi}^{E}_{N,\tau e} \leq 0.05$\,-\,$0.005 \leq
\bar{\xi}^{E}_{N,\tau e} \leq 0.05$\,-\,$50 \leq
\bar{\xi}^{E}_{N,\tau \tau} \leq 100$ it enhances almost
two-two-one order of magnitude. These figures also show that the
inclusion of lepton KK modes causes the BR to increase
considerably.

Finally, we study the effects of the position of the localization
point of the new Higgs doublet on the BR of the considered decays.

Fig. \ref{Zmuealf} represents the parameter $\alpha$ dependence of
BR $\,(Z\rightarrow \mu^{\pm}\, e^{\pm})$ for the Yukawa couplings
$\bar{\xi}^{E}_{N,\tau \mu}=10\,GeV$, $\bar{\xi}^{E}_{N,\tau
e}=0.1\,GeV$, $\rho=0.01$ and $1/R=500\, GeV$. Here the solid
(dashed) line represents the BR without lepton KK modes in a
single extra dimension (with lepton KK modes in a single extra
dimension). Without lepton KK modes, the BR is not sensitive to
the parameter $\alpha$ for the interval $0.1 \leq \alpha \leq 1$.
The inclusion of lepton KK modes makes the BR sensitive to the
parameter $\alpha$ and increasing values of this parameter results
in to decrease the BR almost one order for the interval taken.

Fig. \ref{Ztauealf}-\ref{Ztaumualf} represents the parameter
$\alpha$ dependence of BR$\,(Z\rightarrow \tau^{\pm}\,
e^{\pm})$\,-\,BR$\,(Z\rightarrow \tau^{\pm}\, \mu^{\pm})$ for the
Yukawa couplings $\bar{\xi}^{E}_{N,\tau \tau}=100\,GeV$,
$\bar{\xi}^{E}_{N,\tau e}=0.1\,GeV$\,-\,$\bar{\xi}^{E}_{N,\tau
\mu}=10\,GeV$, $\rho=0.01$ and $1/R=500\, GeV$. Here the solid
(dashed) line represents the BR without lepton KK modes in a
single extra dimension (with lepton KK modes in a single extra
dimension) for both figures. The BR is not sensitive to $\alpha$
for the interval $0.1 \leq \alpha \leq 1$ without lepton KK modes
for both decays. The inclusion of lepton KK modes increases the
sensitivity of the BR to the parameter $\alpha$ and increasing
values of this parameter results in to decrease the BR almost one
order for the considered interval for both decays.

As a summary, the BR $\,(Z\rightarrow \mu^{\pm}\, e^{\pm})$
($\,(Z\rightarrow \tau^{\pm}\, e^{\pm})$, $\,(Z\rightarrow
\tau^{\pm}\, \mu^{\pm})$) enhances up to the values of the order
of $10^{-10}$ ($10^{-8}$, $10^{-5}$) with the inclusion of lepton
KK modes in a single extra dimension. For two extra dimensions,
the numerical value of the BRs are slightly smaller compared to
the single extra dimension case. On the other hand the inclusion
of lepton KK modes makes the BRs sensitive to the parameter
$\alpha$ and increasing values of this parameter results in to
decrease the BR almost one order for the considered interval of
this parameter. With the forthcoming more accurate experimental
measurements of the these decays, the valuable information can be
obtained to detect the effects due to the extra dimensions and the
possible localization of the Higgs doublet. 
\begin{appendix}
\section{The explicit expressions appearing in the text}
Here we present the explicit expressions for $f_{iV}^{(n)}$,
$f_{iA}^{(n)}$, $f_{iM}^{(n)}$ and $f_{iE}^{(n)}$ \cite{EiltZl1l2}
(see eq. (\ref{fVAMEex})):
\begin{eqnarray}
f_{iV}^{(n)}&=& \frac{g}{64\,\pi^2\,cos\,\theta_W} \int_0^1\, dx
\, \frac{1}{m^2_{l_2^+}-m^2_{l_1^-}} \Bigg \{ c_V \,
(m_{l_2^+}+m_{l_1^-})
\nonumber \\
&\Bigg(& (-m_i \, \eta^+_i + m_{l_1^-} (-1+x)\, \eta_i^V)\, ln \,
\frac{L^{self}_ {1,\,h^0}}{\mu^2}+ (m_i \, \eta^+_i - m_{l_2^+}
(-1+x)\, \eta_i^V)\, ln \, \frac{L^{self}_{2,\, h^0}}{\mu^2}
\nonumber \\ &+& (m_i \, \eta^+_i + m_{l_1^-} (-1+x)\, \eta_i^V)\,
ln \, \frac{L^{self}_{1,\, A^0}}{\mu^2} - (m_i \, \eta^+_i +
m_{l_2^+} (-1+x) \,\eta_i^V)\, ln \, \frac{L^{self}_{2,\,
A^0}}{\mu^2} \Bigg) \nonumber \\ &+&
c_A \, (m_{l_2^+}-m_{l_1^-}) \nonumber \\
&\Bigg ( & (-m_i \, \eta^-_i + m_{l_1^-} (-1+x)\, \eta_i^A)\, ln
\, \frac{L^{self}_{1,\, h^0}}{\mu^2} + (m_i \, \eta^-_i +
m_{l_2^+} (-1+x)\, \eta_i^A)\, ln \, \frac{L^{self}_{2,\,
h^0}}{\mu^2} \nonumber \\ &+& (m_i \, \eta^-_i + m_{l_1^-}
(-1+x)\, \eta_i^A)\, ln \, \frac{L^{self}_{1,\, A^0}}{\mu^2} +
(-m_i \, \eta^-_i + m_{l_2^+} (-1+x)\, \eta_i^A)\, ln \,
\frac{L^{self}_{2,\, A^0}}{\mu^2} \Bigg) \Bigg \} \nonumber \\ &-&
\frac{g}{64\,\pi^2\,cos\,\theta_W} \int_0^1\,dx\, \int_0^{1-x} \,
dy \, \Bigg \{ m_i^2 \,(c_A\,
\eta_i^A-c_V\,\eta_i^V)\,(\frac{1}{L^{ver}_{A^0}}+
\frac{1}{L^{ver}_{h^0}}) \nonumber \\ &-& (1-x-y)\,m_i\, \Bigg(
c_A\,  (m_{l_2^+}-m_{l_1^-})\, \eta_i^- \,(\frac{1}{L^{ver}_{h^0}}
- \frac{1}{L^{ver}_{A^0}})+ c_V\, (m_{l_2^+}+m_{l_1^-})\, \eta_i^+
\, (\frac{1}{L^{ver}_{h^0}} + \frac{1}{L^{ver}_{A^0}}) \Bigg)
\nonumber \\ &-& (c_A\, \eta_i^A+c_V\,\eta_i^V) \Bigg (
-2+(q^2\,x\,y+m_{l_1^-}\,m_{l_2^+}\, (-1+x+y)^2)\,
(\frac{1}{L^{ver}_{h^0}} +
\frac{1}{L^{ver}_{A^0}})-ln\,\frac{L^{ver}_{h^0}}{\mu^2}\,
\frac{L^{ver}_{A^0}}{\mu^2} \Bigg ) \nonumber \\ &-&
(m_{l_2^+}+m_{l_1^-})\, (1-x-y)\, \Bigg (
\frac{\eta_i^A\,(x\,m_{l_1^-} +y\,m_{l_2^+})+m_i\,\eta_i^-}
{2\,L^{ver}_{A^0\,h^0}}+\frac{\eta_i^A\,(x\,m_{l_1^-}
+y\,m_{l_2^+})- m_i\,\eta_i^-}{2\,L^{ver}_{h^0\,A^0}} \Bigg )
\nonumber \\ &+& \frac{1}{2}\eta_i^A\,
ln\,\frac{L^{ver}_{A^0\,h^0}}{\mu^2}\,
\frac{L^{ver}_{h^0\,A^0}}{\mu^2}
\Bigg \}\,, \nonumber \\
f_{iA}^{(n)}&=& \frac{-g}{64\,\pi^2\,cos\,\theta_W} \int_0^1\, dx
\, \frac{1}{m^2_{l_2^+}-m^2_{l_1^-}} \Bigg \{ c_V \,
(m_{l_2^+}-m_{l_1^-})
\nonumber \\
&\Bigg(& (m_i \, \eta^-_i + m_{l_1^-} (-1+x)\, \eta_i^A)\, ln \,
\frac{L^{self}_{1,\,A^0}}{\mu^2} + (-m_i \, \eta^-_i + m_{l_2^+}
(-1+x)\, \eta_i^A)\, ln \, \frac{L^{self}_ {2,\,A^0}}{\mu^2}
\nonumber \\ &+& (-m_i \, \eta^-_i + m_{l_1^-} (-1+x)\,
\eta_i^A)\, ln \, \frac{L^{self}_{1,\, h^0}}{\mu^2}+ (m_i \,
\eta^-_i + m_{l_2^+} (-1+x)\, \eta_i^A)\, ln \,
\frac{L^{self}_{2,\,h^0}}{\mu^2} \Bigg) \nonumber \\ &+&
c_A \, (m_{l_2^+}+m_{l_1^-}) \nonumber \\
&\Bigg(& (m_i \, \eta^+_i + m_{l_1^-} (-1+x)\, \eta_i^V)\, ln \,
\frac{L^{self}_{1,\, A^0}}{\mu^2}- (m_i \, \eta^+_i + m_{l_2^+}
(-1+x)\, \eta_i^V)\, ln \, \frac{L^{self}_{2,\,A^0}}{\mu^2}
\nonumber \\ &+& (-m_i \, \eta^+_i + m_{l_1^-} (-1+x)\,
\eta_i^V)\, ln \, \frac{L^{self}_{1,\, h^0}}{\mu^2} + (m_i \,
\eta^+_i - m_{l_2^+} (-1+x)\, \eta_i^V)\, \frac{ln \,
L^{self}_{2,\,h^0}}{\mu^2} \Bigg) \Bigg \} \nonumber \\ &+&
\frac{g}{64\,\pi^2\,cos\,\theta_W} \int_0^1\,dx\, \int_0^{1-x} \,
dy \, \Bigg \{ m_i^2 \,(c_V\,
\eta_i^A-c_A\,\eta_i^V)\,(\frac{1}{L^{ver}_{A^0}}+
\frac{1}{L^{ver}_{h^0}}) \nonumber \\ &-& m_i\, (1-x-y)\, \Bigg(
c_V\, (m_{l_2^+}-m_{l_1^-}) \,\eta_i^- + c_A\,
(m_{l_2^+}+m_{l_1^-})\, \eta_i^+ \Bigg) \,(\frac{1}
{L^{ver}_{h^0}} - \frac{1}{L^{ver}_{A^0}}) \nonumber \\ &+& (c_V\,
\eta_i^A+c_A\,\eta_i^V) \Bigg(-2+(q^2\,x\,y-m_{l_1^-}\,m_{l_2^+}\,
(-1+x+y)^2) (\frac{1}{L^{ver}_{h^0}}+\frac{1}{L^{ver}_{A^0}})-
ln\,\frac{L^{ver}_{h^0}}{\mu^2}\,\frac{L^{ver}_{A^0}}{\mu^2}
\Bigg) \nonumber \\ &-& (m_{l_2^+}-m_{l_1^-})\, (1-x-y)\, \Bigg(
\frac{\eta_i^V\,(x\,m_{l_1^-} -y\,m_{l_2^+})+m_i\,\eta_i^+}
{2\,L^{ver}_{A^0\,h^0}}+ \frac{\eta_i^V\,(x\,m_{l_1^-}
-y\,m_{l_2^+})-m_i\, \eta_i^+}{2\,L^{ver}_{h^0\,A^0}}
\Bigg)\nonumber \\
&-& \frac{1}{2} \eta_i^V\, ln\,\frac{L^{ver}_{A^0\,h^0}}{\mu^2}\,
\frac{L^{ver}_{h^0\,A^0}}{\mu^2}
\Bigg \} \nonumber \,,\\
f_{iM}^{(n)}&=&-\frac{g\, m_W}{64\,\pi^2\,cos\,\theta_W}
\int_0^1\,dx\, \int_0^{1-x} \, dy \, \Bigg \{ \Bigg(
(1-x-y)\,(c_V\, \eta_i^V+c_A\,\eta_i^A)\, (x\,m_{l_1^-}
+y\,m_{l_2^+}) \nonumber
\\ &+& \, m_i\,(c_A\, (x-y)\,\eta_i^-+c_V\,\eta_i^+\,(x+y))\Bigg )
\,\frac{1}{L^{ver}_{h^0}} \nonumber \\ &+& \Bigg( (1-x-y)\, (c_V\,
\eta_i^V+c_A\,\eta_i^A)\, (x\,m_{l_1^-} +y\,m_{l_2^+})
-m_i\,(c_A\, (x-y)\,\eta_i^-+c_V\,\eta_i^+\,(x+y))\Bigg )
\,\frac{1}{L^{ver}_{A^0}} \nonumber \\ &-& (1-x-y) \Bigg
(\frac{\eta_i^A\,(x\,m_{l_1^-} +y\,m_{l_2^+})}{2}\, \Big (
\frac{1}{L^{ver}_{A^0\,h^0}}+\frac{1}{L^{ver}_{h^0\,A^0}} \Big )
+\frac{m_i\,\eta_i^-} {2} \, \Big ( \frac{1}{L^{ver}_{h^0\,A^0}}-
\frac{1}{L^{ver}_{A^0\,h^0}} \Big ) \Bigg ) \Bigg \} \,,\nonumber \\
f_{iE}^{(n)}&=&-\frac{g\, m_W}{64\,\pi^2\, cos\,\theta_W}
\int_0^1\,dx\, \int_0^{1-x} \, dy \, \Bigg \{ \Bigg( (1-x-y)\,\Big
( -(c_V\, \eta_i^A+c_A\,\eta_i^V)\, (x\,m_{l_1^-} -y\, m_{l_2^+})
\Big) \nonumber \\ &-& m_i\, (c_A\,
(x-y)\,\eta_i^++c_V\,\eta_i^-\,(x+y))\Bigg )\,
\frac{1}{L^{ver}_{h^0}} \nonumber \\ &+& \Bigg ( (1-x-y)\,\Big (
-(c_V\, \eta_i^A+c_A\,\eta_i^V)\, (x\,m_{l_1^-} - y\, m_{l_2^+})
\Big ) + m_i\,(c_A\, (x-y)\,\eta_i^++c_V\,\eta_i^-\,(x+y)) \Bigg )
\,\frac{1}{L^{ver}_{A^0}} \nonumber \\&+& (1-x-y)\, \Bigg (
\frac{\eta_i^V}{2}\,(m_{l_1^-}\,x-m_{l_2^+}\, y)\, \, \Big (
\frac{1}{L^{ver}_{A^0\,h^0}}+\frac{1}{L^{ver}_{h^0\,A^0}} \Big )
+\frac{m_i\,\eta_i^+}{2}\, \Big (
\frac{1}{L^{ver}_{A^0\,h^0}}-\frac{1}{L^{ver}_{h^0\,A^0}} \Big )
\Bigg ) \Bigg \}, \label{fVAME}
\end{eqnarray}
where
\begin{eqnarray}
L^{self}_{1,\,h^0}&=& m^{
2}_{h^0}\,(1-x)+(m_i^{(n)2}-m^2_{l_1^-}\,(1-x))\,x
\nonumber \, , \\
L^{self}_{1,\,A^0}&=&L^{self}_{1,\,h^0}(m_{h^0}\rightarrow
m_{A^0})
\nonumber \, , \\
L^{self}_{2,\,h^0}&=&L^{self}_{1,\,h^0}(m_{l_1^-}\rightarrow
m_{l_2^+})
\nonumber \, , \\
L^{self}_{2,\,A^0}&=&L^{self}_{1,\,A^0}(m_{l_1^-}\rightarrow
m_{l_2^+})
\nonumber \, , \\
L^{ver}_{h^0}&=& m^{2}_{h^0}\,(1-x-y)+m_i^{(n)2}\,(x+y)-q^2\,x\,y
\nonumber \, , \\
L^{ver}_{h^0\,A^0}&=&m^{2}_{A^0}\,x+m_i^{(n)2}\,(1-x-y)+(m^{
2}_{h^0}-q^2\,x)\,y
\nonumber \, , \\
L^{ver}_{A^0}&=&L^{ver}_{h^0}(m_{h^0}\rightarrow m_{A^0})
\nonumber \, , \\
L^{ver}_{A^0\,h^0}&=&L^{ver}_{h^0\,A^0}(m_{h^0}\rightarrow
m_{A^0}) \, , \label{Lh0A0}
\end{eqnarray}
and
\begin{eqnarray}
\eta_i^V&=& c_n^2 \,\{ \xi^{E}_{il_1}\xi^{E\,*}_{il_2}+
\xi^{E\,*}_{l_1i} \xi^{E}_{l_2 i} \}\nonumber \, , \\
\eta_i^A&=& c_n^2 \,\{ \xi^{E}_{il_1}\xi^{E\,*}_{il_2}-
\xi^{E\,*}_{l_1i} \xi^{E}_{l_2 i}\} \nonumber \, , \\
\eta_i^+&=& c_n^2 \,\{ \xi^{E\,*}_{l_1i}\xi^{E\,*}_{il_2}+
\xi^{E}_{il_1} \xi^{E}_{l_2 i} \,\} \nonumber \, , \\
\eta_i^-&=& c_n^2 \,\{ \xi^{E\,*}_{l_1i}\xi^{E\,*}_{il_2}-
\xi^{E}_{il_1} \xi^{E}_{l_2 i}\}\, . \label{etaVA}
\end{eqnarray}
The parameters $c_V$ and $c_A$ are $c_A=-\frac{1}{4}$ and
$c_V=\frac{1}{4}-sin^2\,\theta_W$ and the masses $m^{(n)}_{i}$
read $m^{(n)}_{i}=\sqrt{m_i^2+n^2/R^2}$, where $R$ is the
compactification radius. In eq. (\ref{etaVA}) the flavor changing
couplings $\xi ^{E}_{i l_j}$ represent the effective interaction
between the internal lepton $i$, ($i=e,\mu,\tau$) and outgoing
(incoming) $j=1\,(j=2)$ one. The parameter $c_n$ is defined in eq.
(\ref{cevenodd}) for the localization of the new Higgs doublet
around the origin and in eq. (\ref{cevenoddH}) for the
localization of the new Higgs doublet around the point $y_H$ near
to the origin. In the case of two extra dimensions $c_n$ is
replaced by $c_{n,s}$ (see eq. (\ref{cevenodd2})) and the masses
$m^{(n)}_{i}$ are replaced by $m^{(n,s)}_{i}$,
$m^{(n,s)}_{i}=\sqrt{m^2_{i}+m_n^2+m_s^2}$, with $m_n=n/R, \,
m_s=s/R$.

Finally, the couplings $\xi^{E}_{l_ji}$ may be complex in general
and they can be parametrized as
\begin{eqnarray}
\xi^{E}_{l_ij }=|\xi^{E}_{l_ij }|\, e^{i \theta_{ij}} \,\, ,
\label{xi}
\end{eqnarray}
where $i,l_j$ denote the lepton flavors and $\theta_{ij}$ are CP
violating parameters which are the possible sources of the lepton
EDM. However, in the present work we take these couplings real.
\end{appendix}
\newpage
\newpage
\begin{figure}[htb]
\vskip 0.0truein \centering \epsfxsize=6.0in
\leavevmode\epsffile{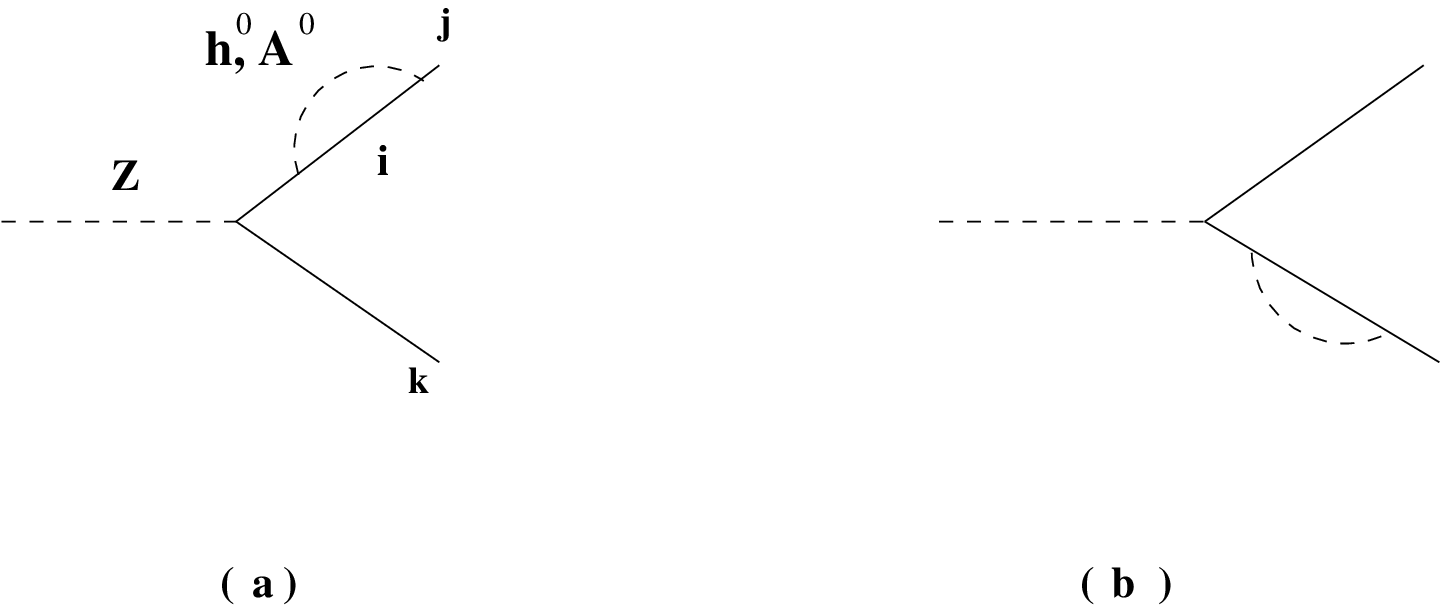} \vskip 0.5truein
\end{figure}
\begin{figure}[htb]
\vskip -0.5truein \centering \epsfxsize=6.0in
\leavevmode\epsffile{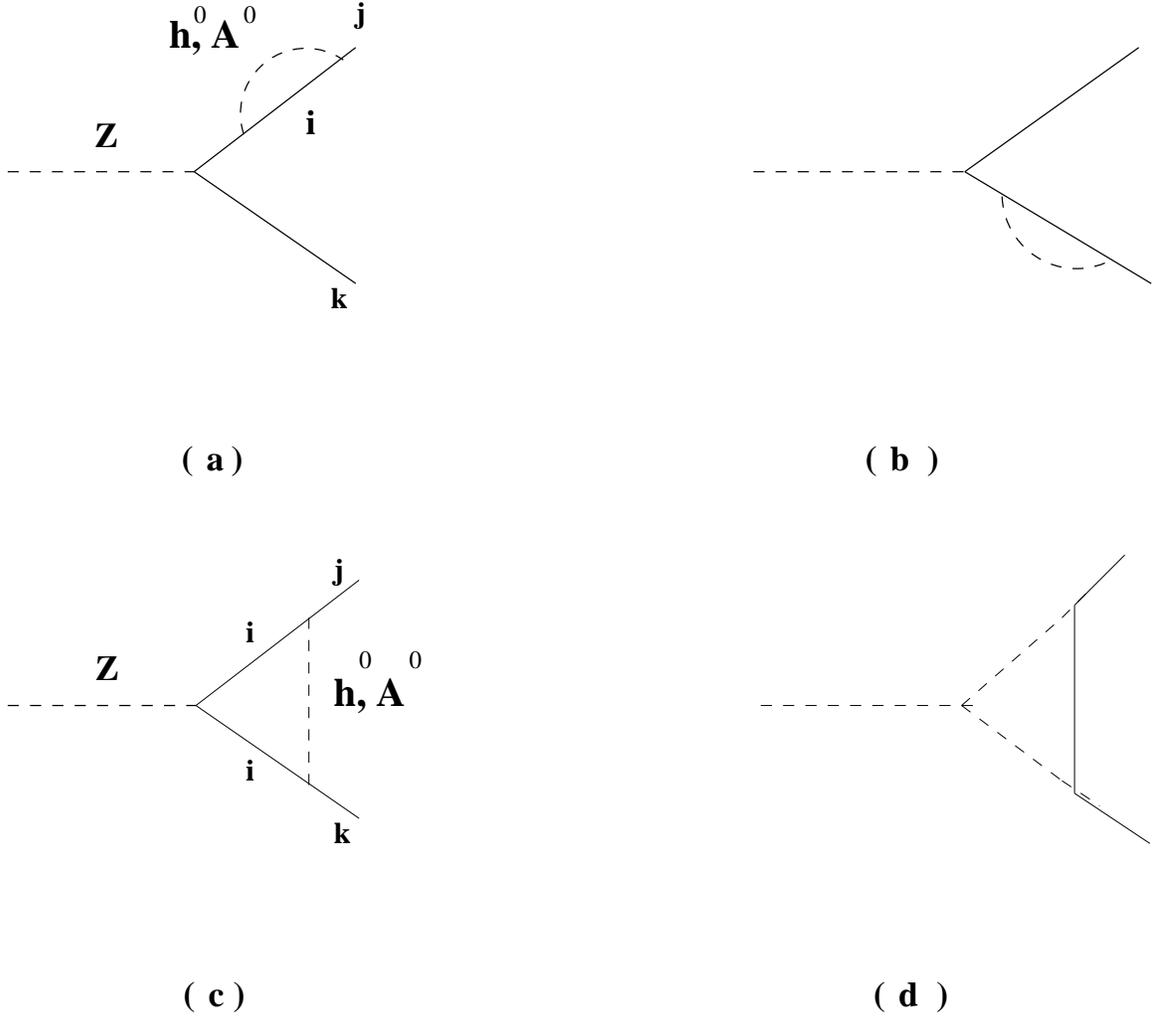} \vskip 0.5truein \caption[]{One
loop diagrams contribute to $Z\rightarrow k^+\,j^-$ decay due to
the neutral Higgs bosons $h_0$ and $A_0$ in the 2HDM. $i$
represents the internal, $j$ ($k$) outgoing (incoming) lepton,
dashed lines the vector field Z, $h_0$ and $A_0$ fields. In 5 (6)
dimensions there exist also the KK modes of lepton and Higgs
fields.} \label{fig1ver}
\end{figure}
\newpage
\begin{figure}[htb]
\vskip -3.0truein \centering \epsfxsize=6.8in
\leavevmode\epsffile{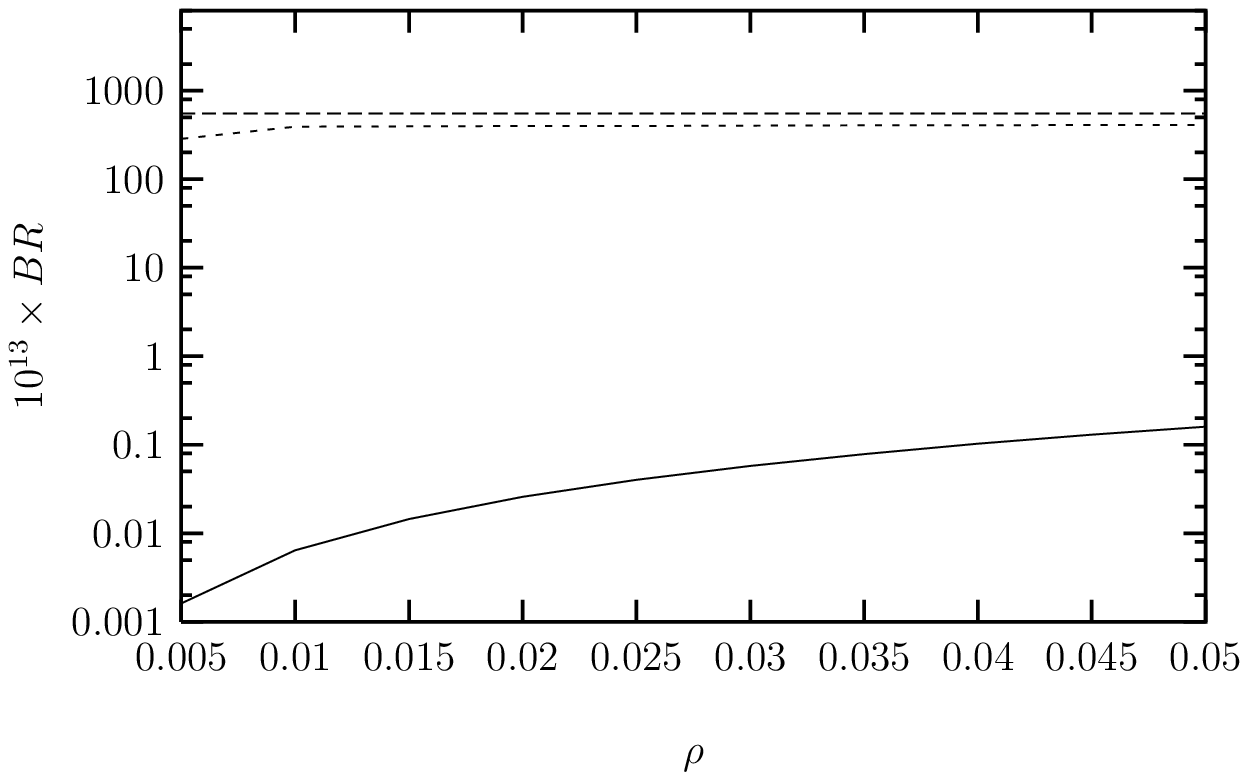} \vskip -3.0truein
\caption[]{BR($\,Z\rightarrow \mu^{\pm}\, e^{\pm}$) with respect
to the parameter $\rho$  for $\bar{\xi}^{D}_{N,\tau e}=0.1\, GeV$,
$\bar{\xi}^{D}_{N,\tau \mu}=10\,GeV$ and $1/R=500\, GeV$. Here the
solid (dashed, small dashed) line represents the BR without lepton
KK modes in a single extra dimension (with lepton KK modes in a
single extra dimension, with lepton KK modes in two extra
dimensions).} \label{Zmuero}
\end{figure}
\begin{figure}[htb]
\vskip -3.0truein \centering \epsfxsize=6.8in
\leavevmode\epsffile{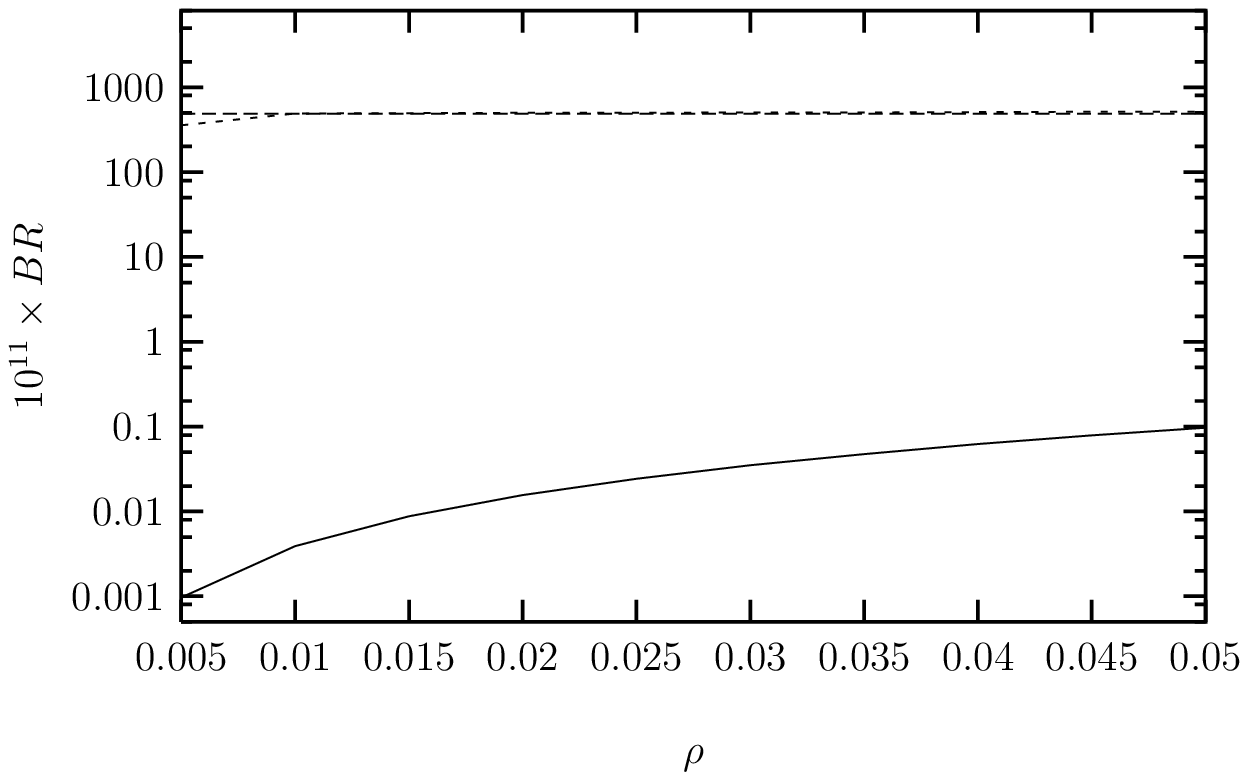} \vskip -3.0truein
\caption[]{BR($\,Z\rightarrow \tau^{\pm}\, e^{\pm}$) with respect
to the parameter $\rho$  for $\bar{\xi}^{D}_{N,\tau e}=0.1\, GeV$,
$\bar{\xi}^{D}_{N,\tau \tau}=100\,GeV$ and $1/R=500\, GeV$. Here
the solid (dashed, small dashed) line represents the BR without
lepton KK modes in a single extra dimension (with lepton KK modes
in a single extra dimension, with lepton KK modes in two extra
dimensions).} \label{Ztauero}
\end{figure}
\begin{figure}[htb]
\vskip -3.0truein \centering \epsfxsize=6.8in
\leavevmode\epsffile{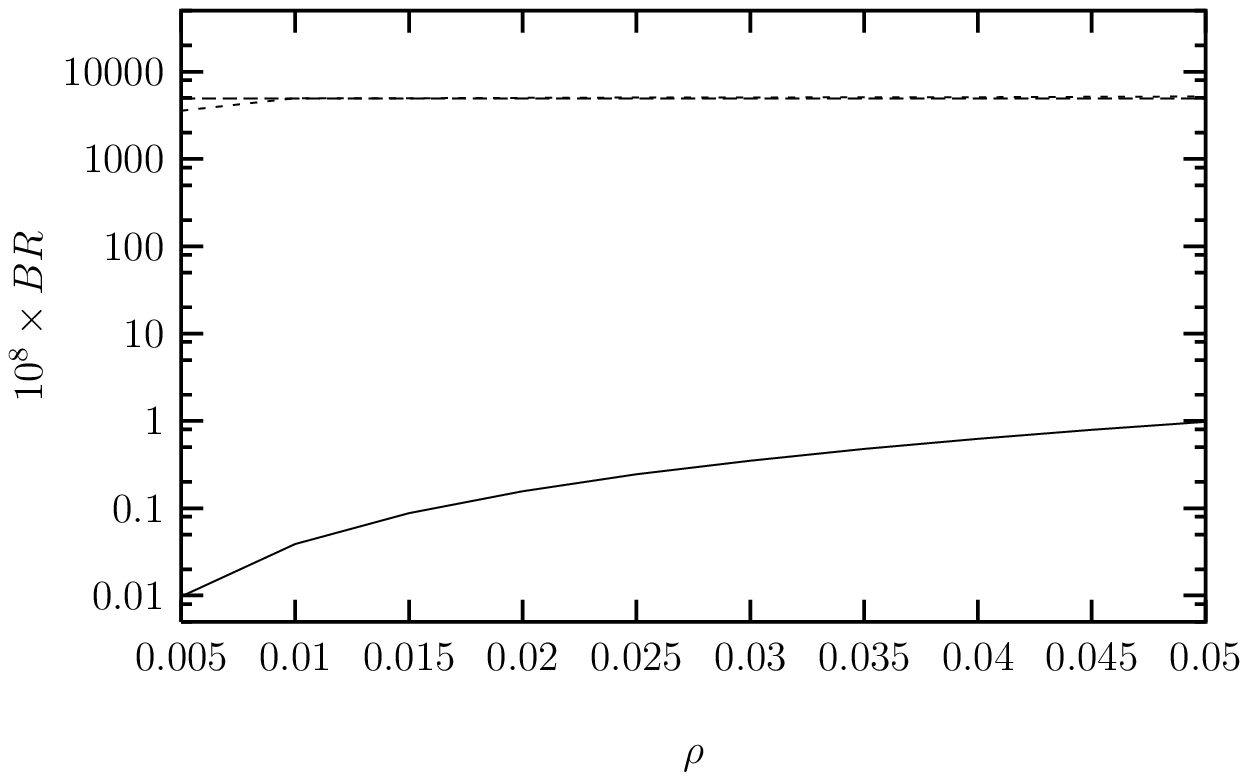} \vskip -3.0truein
\caption[]{BR($\,Z\rightarrow \tau^{\pm}\, \mu^{\pm}$) with
respect to the parameter $\rho$  for $\bar{\xi}^{D}_{N,\tau
\mu}=10\, GeV$, $\bar{\xi}^{D}_{N,\tau \tau}=100\,GeV$ and
$1/R=500\, GeV$. Here the solid (dashed, small dashed) line
represents the BR without lepton KK modes in a single extra
dimension (with lepton KK modes in a single extra dimension, with
lepton KK modes in two extra dimensions).} \label{Ztaumuro}
\end{figure}
\begin{figure}[htb]
\vskip -3.0truein \centering \epsfxsize=6.8in
\leavevmode\epsffile{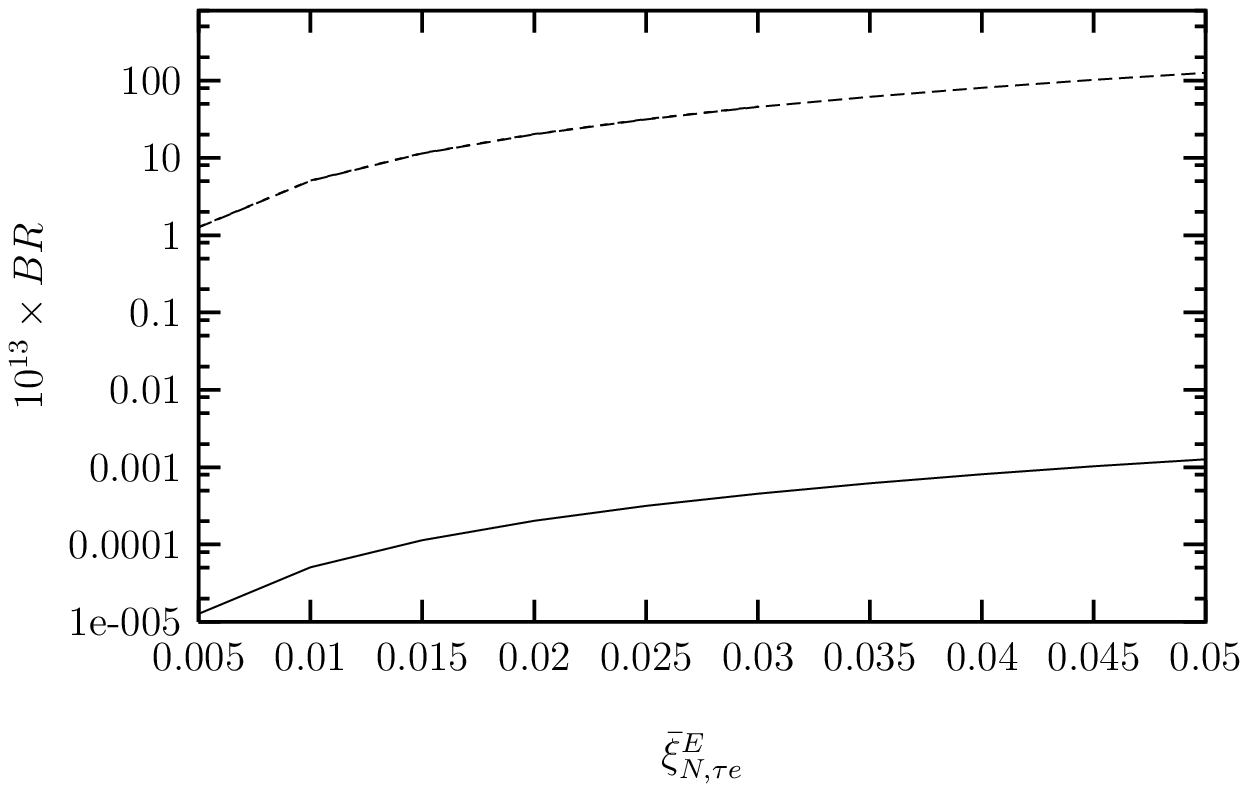} \vskip -3.0truein
\caption[]{$\,Z\rightarrow \mu^{\pm}\, e^{\pm}$ with respect to
$\bar{\xi}^{E}_{N,\tau e}$ for $\bar{\xi}^{D}_{N,\tau \mu}=10\,
GeV$, $\rho=0.01$ and $1/R=500\, GeV$. Here the solid (dashed,
small dashed) line represents the BR without lepton KK modes in a
single extra dimension (with lepton KK modes in a single extra
dimension, with lepton KK modes in two extra dimensions).}
\label{Zmueksi}
\end{figure}
\begin{figure}[htb]
\vskip -3.0truein \centering \epsfxsize=6.8in
\leavevmode\epsffile{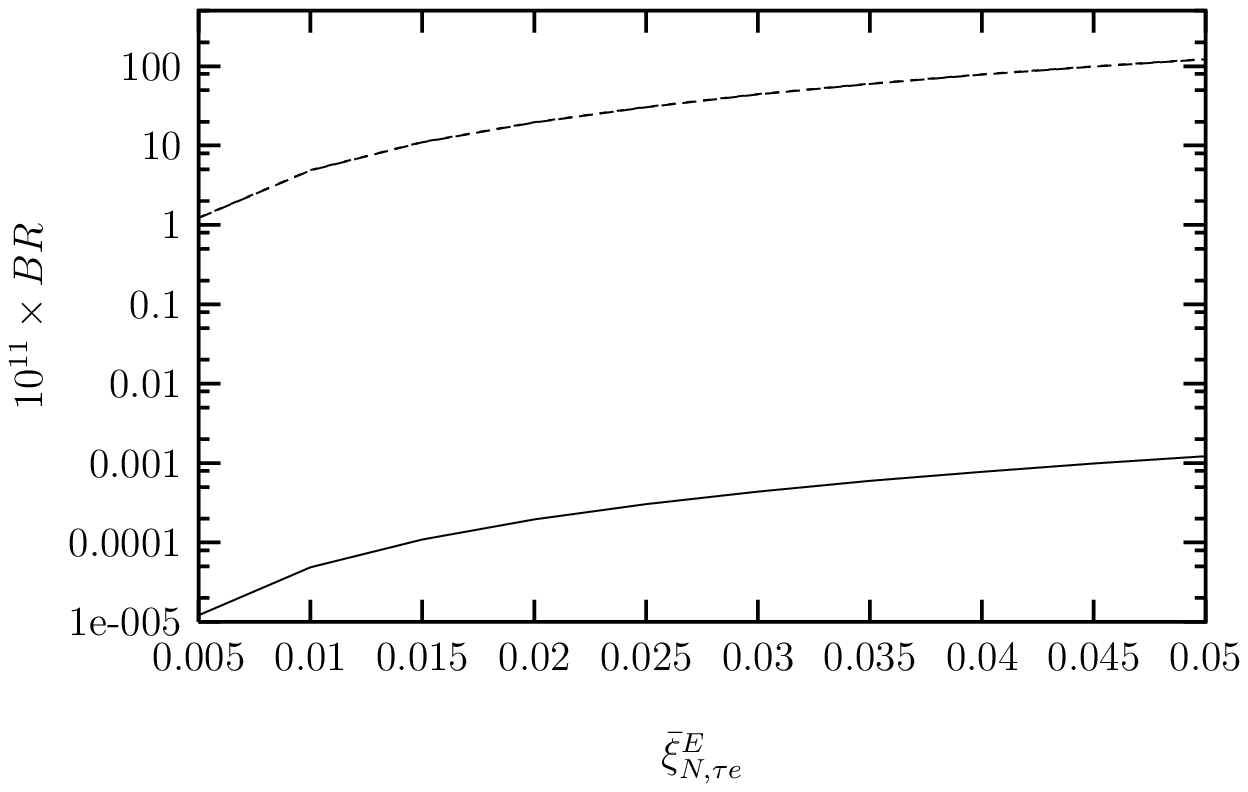} \vskip -3.0truein
\caption[]{$\,Z\rightarrow \tau^{\pm}\, e^{\pm}$ with respect to
$\bar{\xi}^{E}_{N,\tau e}$ for $\bar{\xi}^{D}_{N,\tau \tau}=100\,
GeV$, $\rho=0.01$ and $1/R=500\, GeV$. Here the solid (dashed,
small dashed) line represents the BR without lepton KK modes in a
single extra dimension (with lepton KK modes in a single extra
dimension, with lepton KK modes in two extra dimensions).}
\label{Ztaueksi}
\end{figure}
\begin{figure}[htb]
\vskip -3.0truein \centering \epsfxsize=6.8in
\leavevmode\epsffile{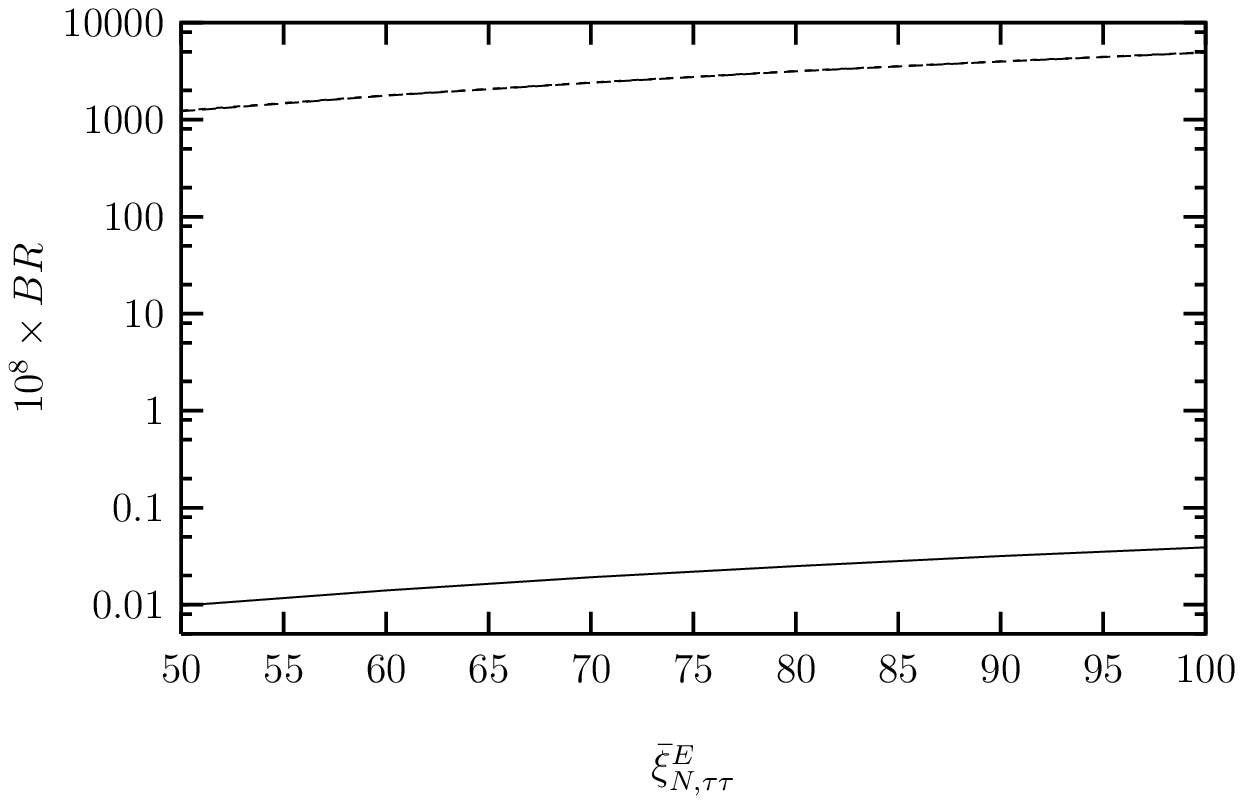} \vskip -3.0truein
\caption[]{$\,Z\rightarrow \tau^{\pm}\, \mu^{\pm}$ with respect to
$\bar{\xi}^{E}_{N,\tau \tau}$ for $\bar{\xi}^{D}_{N,\tau \mu}=10\,
GeV$, $\rho=0.01$ and $1/R=500\, GeV$. Here the solid (dashed,
small dashed) line represents the BR without lepton KK modes in a
single extra dimension (with lepton KK modes in a single extra
dimension, with lepton KK modes in two extra dimensions).}
\label{Ztaumuksi}
\end{figure}
\begin{figure}[htb]
\vskip -3.0truein \centering \epsfxsize=6.8in
\leavevmode\epsffile{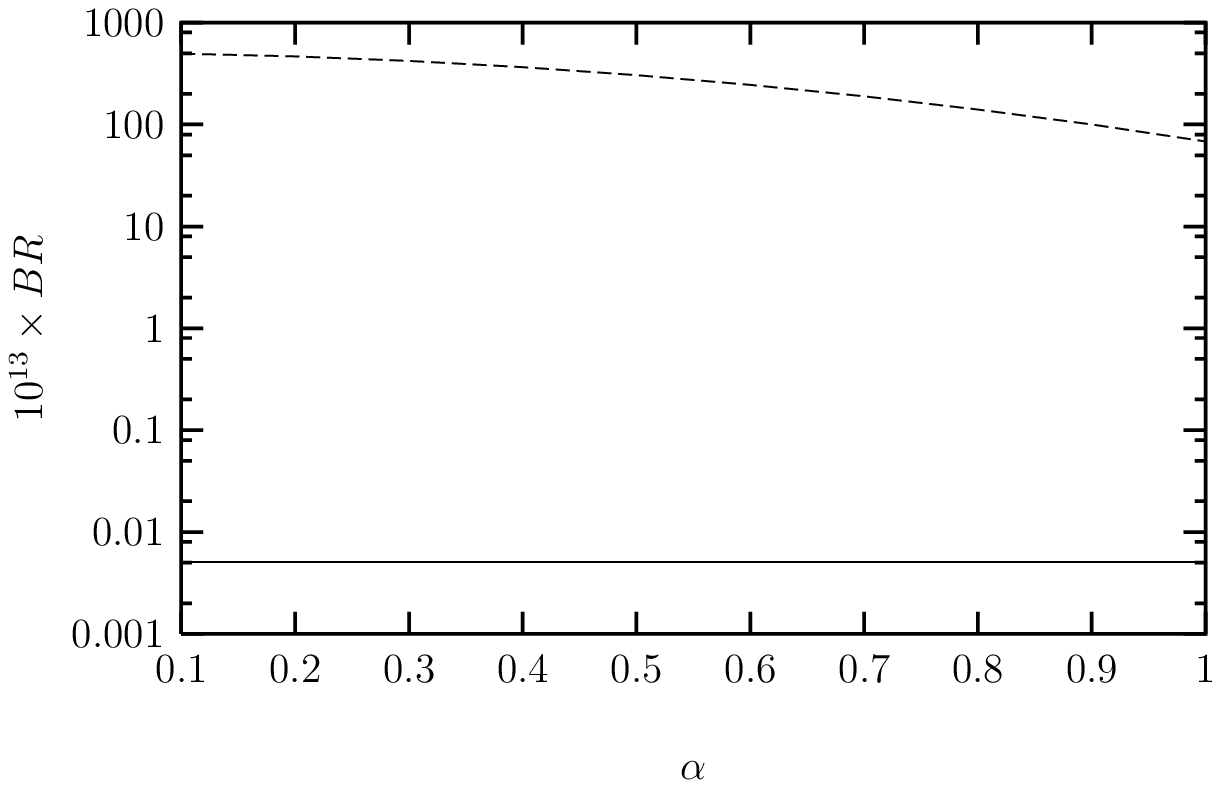} \vskip -3.0truein
\caption[]{$\,Z\rightarrow \mu^{\pm}\, e^{\pm}$ with respect to
the parameter $\alpha$ for $\bar{\xi}^{E}_{N,\tau \mu}=10\,GeV$,
$\bar{\xi}^{E}_{N,\tau e}=0.1\,GeV$, $\rho=0.01$ and $1/R=500\,
GeV$. Here the solid (dashed) line represents the BR without
lepton KK modes in a single extra dimension (with lepton KK modes
in a single extra dimension).} \label{Zmuealf}
\end{figure}
\begin{figure}[htb]
\vskip -3.0truein \centering \epsfxsize=6.8in
\leavevmode\epsffile{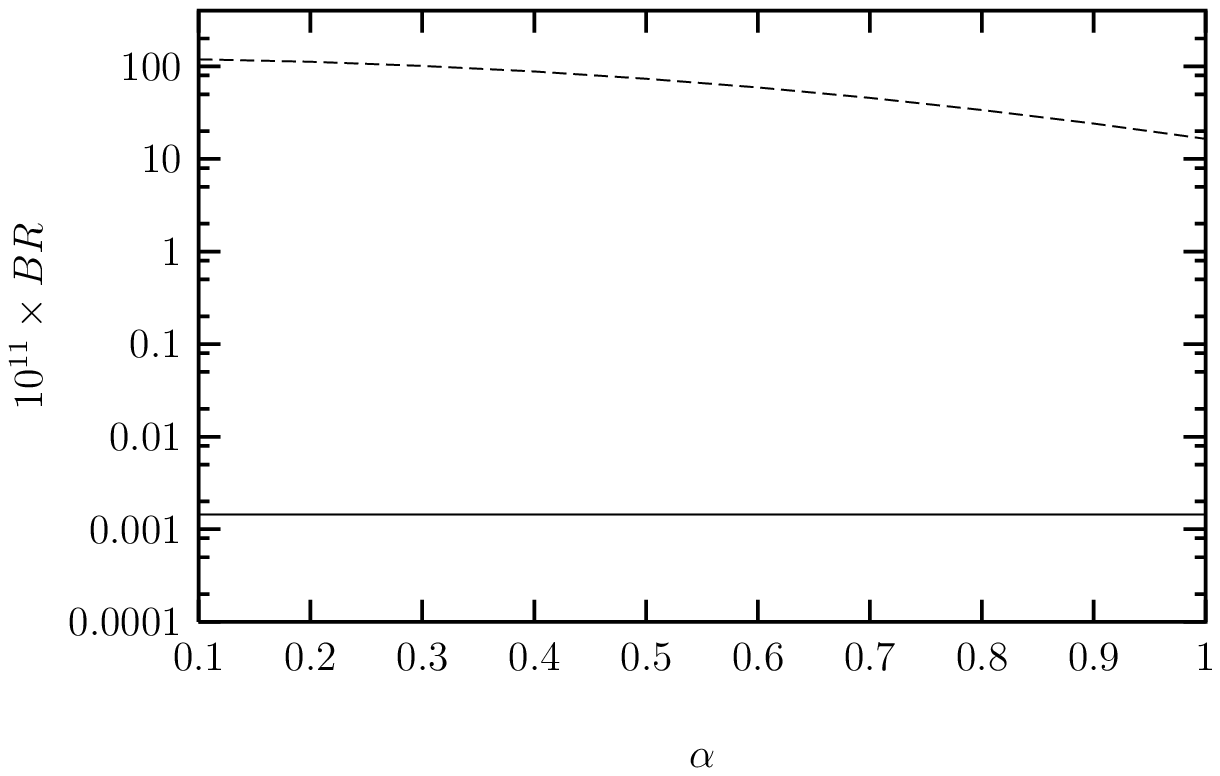} \vskip -3.0truein
\caption[]{$\,Z\rightarrow \tau^{\pm}\, e^{\pm}$ with respect to
the parameter $\alpha$ for $\bar{\xi}^{E}_{N,\tau \tau}=100\,GeV$,
$\bar{\xi}^{E}_{N,\tau e}=0.1\,GeV$, $\rho=0.01$ and $1/R=500\,
GeV$. Here the solid (dashed) line represents the BR without
lepton KK modes in a single extra dimension (with lepton KK modes
in a single extra dimension).} \label{Ztauealf}
\end{figure}
\begin{figure}[htb]
\vskip -3.0truein \centering \epsfxsize=6.8in
\leavevmode\epsffile{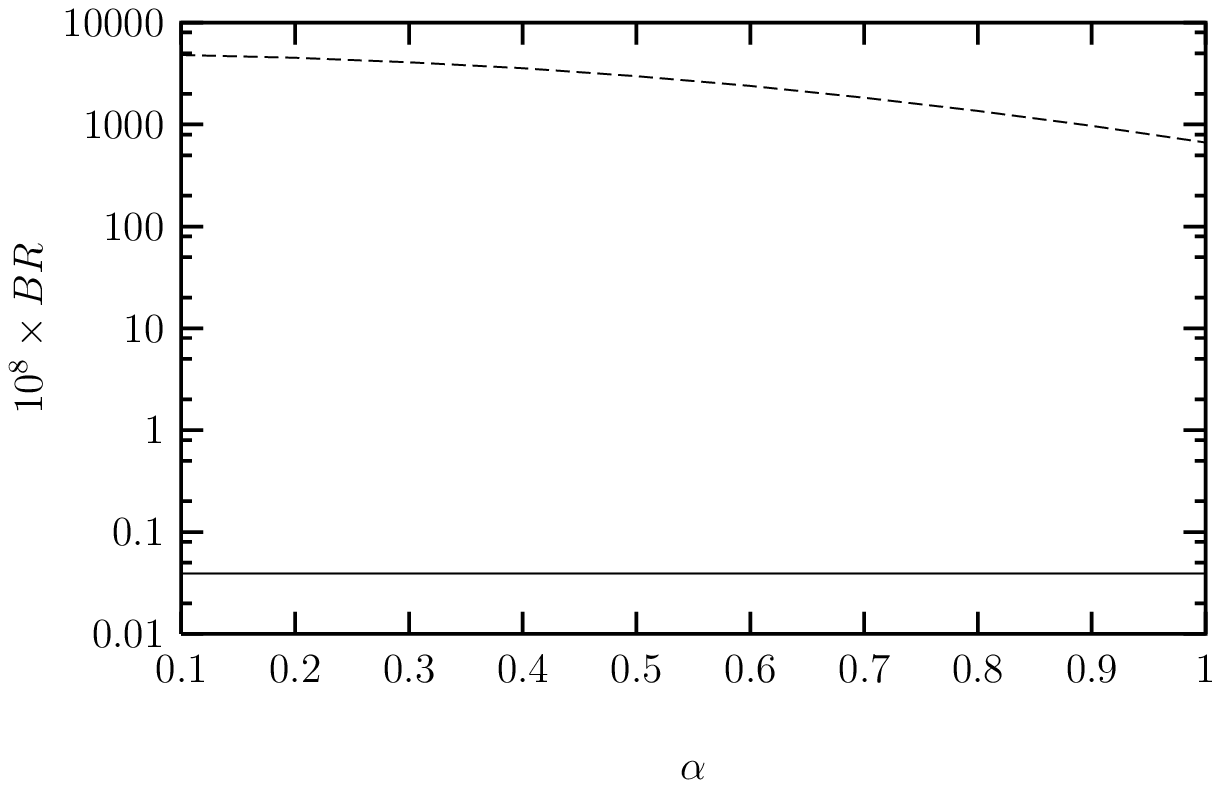} \vskip -3.0truein
\caption[]{$\,Z\rightarrow \tau^{\pm}\, \mu^{\pm}$ with respect to
the parameter $\alpha$ for $\bar{\xi}^{E}_{N,\tau \tau}=100\,GeV$,
$\bar{\xi}^{E}_{N,\tau \mu}=10\,GeV$, $\rho=0.01$ and $1/R=500\,
GeV$. Here the solid (dashed) line represents the BR without
lepton KK modes in a single extra dimension (with lepton KK modes
in a single extra dimension).} \label{Ztaumualf}
\end{figure}

\begin{thebibliography}{1}
%
\bibitem{Riemann}
T. Riemann and G. Mann, ``Nondiagonal {$Z$} decay: ${Z} \to e
\mu$'', in {\it Proc. of the Int. Conf. Neutrino'82, 14-19 June
1982, Balatonf{\"u}red, Hungary} (A. Frenkel and E. Jenik, eds.),
vol. II, pp. 58-, Budapest, 1982, scanned copy at
http://www.ifh.de/$\sim$riemann, G. Mann and T. Riemann, {\it
Annalen Phys.} {\bf 40} (1984) 334.
%
\bibitem{Illana}
J. I. Illana, M. Jack and T. Riemann, {\it hep-ph/0001273 }
(2000), J. I. Illana, and T. Riemann, {\it Phys. Rev.} {\bf D63 }
053004 (2001).
%
\bibitem{PartData}
{Particle Data Group} Collaboration, {C. Caso} {\it et al.}, {\it
Eur. Phys. J.} {\bf C3} (1998) 1.
%
\bibitem{Opal}
{OPAL} Collaboration, R. Akers {\it et al.}, {\it Z. Phys.} {\bf
C67} (1995) 555.
%
\bibitem{L3}
{L3} Collaboration, O. Adriani {\it et al.}, {\it Phys. Lett.}
{\bf B316} (1993) 427.
%
\bibitem{Delphi}
{DELPHI} Collaboration, P. Abreu {\it et al.}, {\it Z. Phys.} {\bf
C73} (1997) 243.
%
\bibitem{Wilson}
G. Wilson, ``Neutrino oscillations: are lepton-flavor violating Z
decays observable with the CDR detector?'' and ``Update on
experimental aspects of lepton-flavour violation'', talks at
DESY-ECFA LC Workshops held at Frascati, Nov 1998 and at Oxford,
March 1999,
transparencies obtainable at \\
http://wwwsis.lnf.infn.it/talkshow/ and at \\
http://hepnts1.rl.ac.uk/ECFA\_DESY\_OXFORD/scans/0025\_wilson.pdf.
%
\bibitem{Ghosal} A. Ghosal, Y. Koide and H. Fusaoka,
{\it Phys.Rev.} {\bf D64} (2001) 053012.
%
\bibitem{EiltZl1l2} E. O. Iltan, I. Turan,
{\it Phys.Rev.} {\bf D65} (2002) 013001.
%
\bibitem{EiltZl1l2Extr} E. O. Iltan, {\it Eur.
Phys. J} {\bf C41} (2005) 233; E. O. Iltan, hep-ph/0507213.
%
\bibitem{Masip} J. Illiana, M. Masip, {\it Phys. Rev.} {\bf D67} (2003)
035004.
%
\bibitem{Cao} J. Cao, Z. Xiong, J. Mi. Yang, {\it Eur. Phys. J.}
{\bf C32}(2004) 245.
%
\bibitem{Yue} C. Yue, W. Wang, F. Zhang,
{\it J. Phys.} {\bf G30} (2004) 1065.
%
\bibitem{Perez} M. A .Perez, G. T. Velasco, J. J. Toscano,
{\it Int.J.Mod.Phys.} {\bf A19} (2004) 159;  A. Flores-Tlalpa,
J.M. Hernandez, G. Tavares-Velasco, J.J. Toscano, {\it Phys. Rev.}
{\bf D65} (2002) 073010, R. Hawkings and K. M{\"o}nig, {\it Eur.
Phys. J. direct} {\bf C8} (1999) 1; V. Ganapathi, T. Weiler, E.
Laermann, I. Schmitt, and P. Zerwas, {\it Phys.  Rev.} {\bf D27}
(1983) 579, M. Clements, C. Footman, A. Kronfeld, S. Narasimhan,
and D. Photiadis, {\it Phys. Rev.} {\bf D27} (1983) 570.
%
\bibitem{Pontecorvo}
B. Pontecorvo, {\it Zh. Eksp. Teor. Fiz.} {\bf 33} (1957) 549, Z.
Maki, M. Nakagawa, and S. Sakata, {\it Prog. Theor. Phys.} {\bf
28} (1962) 870, B. Pontecorvo, {\it Sov. Phys. JETP} {\bf 26}
(1968) 984.
%
\bibitem{Kaluza} Th. Kaluza, Sitzungober, Preuss. Akad. Wiss.
Berlin, p. 966 (1921); O. Klein, Z. Phys. 37 (1926) 895.
%
\bibitem{Dvali} I. Antoniadis, {\it et al.}, {\it Phys. Lett.}  {\bf B436}
(1998) 257; I. Antoniadis, S. Dimopoulos, G. Dvali, Nucl. Phys.
{\bf B516} (1998) 70.
%
\bibitem{ArkaniHamed}N. Arkani-Hamed, S. Dimopoulos and  G. Dvali,
{\it Phys. Lett.} {\bf B429} (1998) 263;
%
\bibitem{Arkani} N. Arkani-Hamed, S. Dimopoulos, and G. R. Dvali,
{\it Phys. Rev.} {\bf D59} 086004 (1999).
%
\bibitem{Antoniadis1} I. Antoniadis,{\it Phys. Lett.} {\bf B246},
377 (1990); I.Antoniadis and K. Benakli, {\it Phys. Lett.} {\bf
B326}, 69 (1994).
%
\bibitem{Antoniadis2}
I. Antoniadis, K. Benakli, and M. Quiros, {\it Phys. Lett.} {\bf
B331}, 313 (1994); A. Pomarol and M. Quiros, {\it Phys. Lett.}
{\bf B438}, 255 (1998); I. Antoniadis, K. Benakli, and M. Quiros,
{\it Phys. Lett.} {\bf B460}, 176 (1999); P. Nath, Y. Yamada, and
M. Yamaguchi, {\it Phys. Lett.} {\bf B466}, 100 (1999); M. Masip
and A. Pomarol, {\it Phys. Rev.} {\bf D60}, 096005 (1999) A.
Delgado, A. Pomarol, and M. Quiros, {\it JHEP} {\bf 01}, 030
(2000); P. Nath and M. Yamaguchi, {\it Phys. Rev.} {\bf D60},
116004 (1999); A. Muck, A. Pilaftsis, and R. Ruckl, {\it Phys.
Rev.} {\bf D65}, 085037 (2002); A. Muck, A. Pilaftsis, and R.
Ruckl, hep-ph/0203032; C. D. Carone, {\it Phys. Rev.} {\bf D61},
015008 (2000).
%
\bibitem{Rizzo} T. G. Rizzo and J. D. Wells, {\it Phys. Rev.}
{\bf D61}, 016007 (2000);
%
\bibitem{Antoniadis3}I. Antoniadis, C. Munoz, M. Quiros,
{\it Nucl. Phys} {\bf B397} 515 (1993);
%
\bibitem{Appelquist} T. Appelquist, H.-C. Cheng, B. A. Dobrescu,
 {\it Phys. Rev.} {\bf D64} 035002 (2001).
%
\bibitem{Papavassiliou}
J. Papavassiliou and A. Santamaria, {\it Phys. Rev.} {\bf D63},
016002 (2001).
%
\bibitem{Chakraverty}
D. Chakraverty, K. Huitu, and A. Kundu, Phys.Lett. B558 (2003)
173-181; A. J. Buras, M. Spranger, and A. Weiler, Nucl.Phys. B660
(2003) 225.
%
\bibitem{Agashe} K. Agashe, N.G. Deshpande, G.-H. Wu, {\it Phys. Lett.}
{\bf 514} 309 (2001).
%
\bibitem{Dienes} K. R. Dienes, E. Dudas, T. Gherghetta
9811428, Q. H. Shrihari, S. Gopalakrishna, C. P. Yuan, 0312339
%
\bibitem{Agulia} F. Agulia, M. P. Victoria, J. Santiago,
{\it JHEP} {\bf 0302} 051, (2003); {\it Acta Phys.Polon.} {\bf
B34}, 5511 (2003).
%
\bibitem{iltanEDM} E. O. Iltan, hep-ph/0401229; {\it JHEP} {\bf 0402} 065,
(2004); {\it JHEP} {\bf 0408} 020, (2004); {\it JHEP} {\bf 0404}
018, (2004); {\it Mod. Phys. Lett.} {\bf A20} 1845, (2005).
%
\bibitem{Lam} C. S. Lam, hep-ph/0302227, 2003; C. A. Scrucca, M. Serona,
L. Silvestrini, {\it Nucl.Phys.} {\bf B669} 128, (2003); M. Gozdz,
W. A. Kaminsk, {\it Phys. Rev.} {\bf D68} 057901, (2003) ; C.
Biggio, et.al, {\it Nucl.Phys.} {\bf B677} 451, (2004); M. Carena,
et.al, {\it Phys. Rev.} {\bf D68} 035010, (2003) ; A. J. Buras,
et. al., {\it Nucl.Phys.} {\bf B678} (2004) 455; T. G. Rizzo, {\it
JHEP} {\bf 0308} 051 (2003), A. J. Buras, hep-ph/0307202, 2003; S.
Matsuda, S. Seki, hep-ph/0307361, 2003; R. N. Mohapatra, {\it
Phys. Rev.} {\bf D68} 116001, (2003); B. Lillie, {\it JHEP} {\bf
0312} 030 (2003); A.A Arkhipov, hep-ph/0309327 2003; F.Feruglia,
{\it Eur.Phys.J.} {\bf C33} (2004) S114.
%
\bibitem{Mirabelli}
E. A. Mirabelli, Schmaltz, {\it Phys. Rev.} {\bf D61} (2000)
113011.
%
\bibitem{Changg} W. F. Chang, I. L. Ho and J. N. Ng, {\it Phys. Rev.}
{\bf D66} 076004 (2002).
%
\bibitem{Branco}
G. C. Branco,A. Gouvea, M. N. Rebelo,  {\it Phys. Lett.} {\bf
B506} (2001) 115.
%
\bibitem{Chang2}
W. F. Cang, J. N. Ng, {\it JHEP} {\bf 0212} (2002) 077.
%
\bibitem{Hewett} B. Lillie, J. L. Hewett, {\it Phys. Rev.} {\bf 68}
(2003) 116002.
%
\bibitem{Perez2}
Y. Grossman and G. Perez, {\it Phys. Rev.} {\bf D67} (2003)
015011; {\it Pramana} {\bf 62} (2004) 733.
%
\bibitem{IltanEDMSplit} E. O. Iltan, {\it Eur.Phys.J.} {\bf C44} (2005)
411.
%
\bibitem{Grossman}
Y. Grossman, {\it Int. J. Mod. Phys.} {\bf A15} (2000) 2419; D. E.
Kaplan and T. M. Tait, {\it JHEP} {\bf 0111} (2001) 051; G.
Barenboim,  {\it et. al.}, {\it Phys. Rev.} {\bf D64} (2001)
073005; W. F. Chang, I. L. Ho and J. N. Ng, {\it Phys. Rev.} {\bf
D66} (2002) 076004; W. Skiba and D. Smith, {\it Phys. Rev.} {\bf
D65} (2002) 095002; Y. Grossman, R. Harnik, G. Perez, M. D.
Schwartz and Z. Surujon, {\it Phys. Rev.} {\bf D71} (2005) 056007;
P. Dey, G. Bhattacharya, {\it Phys. Rev.} {\bf D70} (2004) 116012;
Y. Nagatani, G. Perez, {\it JHEP} {\bf 0502} (2005) 068; R.
Harnik, G. Perez, M. D. Schwartz, Y. Shirman, {\it JHEP} {\bf
0503} (2005) 068.
%
\bibitem{Surujon} Z. Surujon, {\it Phys.Rev.} {\bf D73} (2006) 016008.
%
\bibitem{IltanLFVSplitFat} E. O. Iltan, hep-ph/0509096, (2005).
%
\bibitem{iltSplitHiggsLocal} E. O. Iltan, hep-ph/0511241.
%
\bibitem{LocalNewHiggsLFV} E. O. Iltan, hep-ph/hep-ph/0602170.
%
\bibitem{Sher} T. P. Cheng and M. Sher, {\it Phy. Rev.} {\bf D35} (1987) 3383.
%
%
\end{thebibliography}
\end{document}